\documentstyle[prb,twocolumn,aps,eqsecnum]{revtex}
\begin{document}

\twocolumn[\hsize\textwidth\columnwidth\hsize\csname
@twocolumnfalse\endcsname
\title{ Renormalization-group theory for rotating ${\bf ^4}$He near the 
superfluid transition }

\author{ Rudolf Haussmann }

\address{ Sektion Physik, Universit\"at M\"unchen, Theresienstrasse 37, 
D-80333 M\"unchen, Germany }

\date{ submitted to Physical Review B, February 11, 1999, accepted }

\maketitle

\begin{abstract}
The influence of a uniform rotation with frequency $\Omega$ on the critical
behavior of liquid $^4$He near $T_\lambda$ is investigated. We apply our 
recently developed approach $\lbrack$R.\ Haussmann, submitted to
Phys.\ Rev.\ B$\rbrack$ which is a renormalization-group theory based on 
model~{\it F} starting with the calculation of the Green's function in 
Hartree approximation. We calculate the specific heat $C_\Omega(T,\Omega)$, 
the correlation length $\xi(T,\Omega)$, and the thermal-resistivity tensor 
$\rho_{\rm T}(T,\Omega)$ as functions of the temperature $T$ for fixed
values of the rotation frequency $\Omega$. For nonzero $\Omega$ we find that 
all physical quantities are smooth near $T_\lambda$ so that the superfluid
transition is a smooth crossover. We define a frequency-dependent transition
temperature $T_\lambda(\Omega)$ by the maximum of the specific heat and 
predict the shift $T_\lambda(\Omega)-T_\lambda= - 1.2 \, T_\lambda (2m_4 
\xi_0^2 \Omega/\hbar)^{1/2\nu}$. For $T<T_\lambda(\Omega)$ we find mutual
friction between the superfluid and the normal-fluid component caused 
implicitly by the motion of vortex lines and calculate the Vinen coefficients 
$B$ and $B^\prime$.
\end{abstract}

\vskip2pc]

\section{Introduction} \label{S01}
Liquid $^4$He in a uniformly rotating container is influenced by the rotation 
frequency ${\bf\Omega}$ in two ways. First of all, a rotational flow is
created which implies quantized vortices in the superfluid state for 
temperatures below $T_\lambda$. Secondly, the centrifugal forces of the
rotation cause a spatially dependent pressure $P({\bf r})$ which implies an
inhomogeneity in the system. While usually the second influence is negligible,
the rotational flow and the occurrence of vortices plays an essential role
\cite{DO}.

In this paper we consider the critical behavior of uniformly rotating liquid
$^4$He in $d=3$ dimensions for temperatures near $T_\lambda$ and present a 
renormalization-group theory based on model {\it F\,} of Halperin, Hohenberg 
and Siggia \cite{HH}. We will show that the rotation frequency ${\bf \Omega}$
is an external perturbation which drives the system away from criticality. 
According to Feynman \cite{F1} in uniformly rotating superfluid $^4$He 
straight vortex lines are present which are parallel to the rotation axis 
and which are uniformly distributed in the helium. The circulation around 
a vortex line is quantized by $2\pi\hbar/m_4$ where $m_4$ is the mass of a 
$^4$He atom. Consequently, the density of vortex lines ${\bf L}$ with respect 
to the area perpendicular to the rotation axis is directly related to the 
rotation frequency by ${\bf L}=(m_4/\pi\hbar){\bf\Omega}$. The absolute value 
$L=(m_4/\pi\hbar)\Omega$ is the total length of the vortex lines in a unit 
volume and hence is a measure of how many vortices are present in the helium.

\begin{figure}[t]
\vspace*{6.9cm}
\includegraphics{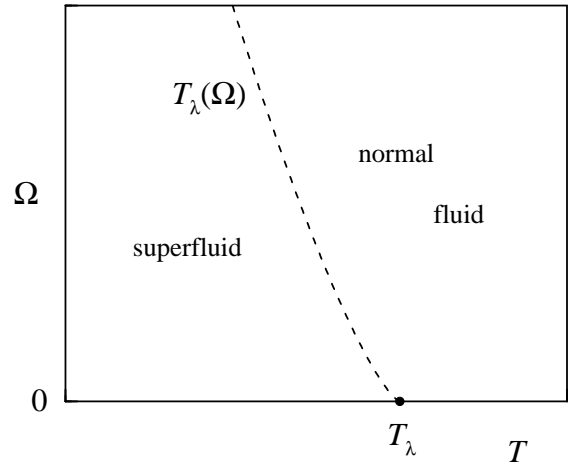}
\caption{The phase diagram of uniformly rotating liquid $^4$He. The dashed
line represents the transition temperature $T_\lambda(\Omega)$ which 
separates the superfluid and the normal-fluid phase. The critical point is 
located at $(T,\Omega)=(T_\lambda,0)$.}
\label{Fig01}
\end{figure}

Close to the superfluid transition the critical fluctuations are very large.
The correlation length $\xi$ increases strongly if the temperature $T$ 
approaches $T_\lambda$. However, for nonzero $\Omega$ the mean distance 
between the vortex lines $L^{-1/2}=(\pi\hbar/m_4\Omega)^{1/2}$ is another
characteristic length of the system. We will show that the correlation length
$\xi$ is bounded by this characteristic length according to $\xi < L^{-1/2}$. 
Consequently, the thermodynamic quantities like the specific heat are smooth 
functions of temperature near $T_\lambda$. As a remnant of the critical 
singularities the quantities exhibit a maximum or an inflection point
at a temperature $T_\lambda(\Omega)$ which is located below $T_\lambda$. We
will find a phase diagram for uniformly rotating $^4$He which is shown 
qualitatively in Fig.\ \ref{Fig01}. The dashed line represents the transition 
temperature $T_\lambda(\Omega)$ between the superfluid and the normal-fluid 
state. This temperature is not sharply defined, because for nonzero $\Omega$
the thermodynamic quantities are nonsingular. The transition is smooth and
shifted to lower temperatures by $\Delta T_\lambda(\Omega) = T_\lambda(\Omega)
-T_\lambda$. Nevertheless, $(T,\Omega)=(T_\lambda,0)$ is the critical point 
in the phase diagram, for which the correlation length $\xi$ diverges and the
thermodynamic quantities are singular. 

The macroscopic quantum coherence of the superfluid component is described by
the complex order-parameter field $\psi({\bf r},t)$. For $T\gtrsim T_\lambda
(\Omega)$ in the normal-fluid phase the order-parameter field $\psi$ 
fluctuates around the average value $\langle\psi\rangle =0$. On the other 
hand, for $T\lesssim T_\lambda(\Omega)$ in the superfluid phase the order
parameter $\psi=\eta\, e^{i\varphi}$ is nonzero. The amplitude $\eta$ 
fluctuates around a nonzero average value. However, the vortex lines in the
uniformly rotating system imply a strong spatial variation of the phase 
$\varphi$ over all values between $0$ and $2\pi$. Since the vortices move
due to fluctuations, the phase $\varphi$ fluctuates strongly so that 
eventually the average order parameter is $\langle\psi\rangle=0$ also in the
superfluid state. 

Conventional field-theoretic methods are not applicable in the present case
because in the superfluid state symmetry breaking with a nonzero average
order parameter $\langle\psi\rangle$ is required for the construction of the 
perturbation theory. However, recently we developed an approach \cite{H1}
which can handle $\langle\psi\rangle=0$ also in the superfluid state and
which includes the effects of vortices in an indirect way. This approach 
starts with the calculation of the order-parameter Green's function in 
self-consistent Hartree approximation and includes the effects of critical 
fluctuations by renormalization and application of the renormalization-group
(RG) theory. In Sec.\ \ref{S02} we apply the approach to uniformly rotating 
$^4$He. We describe the necessary modifications to include the rotation with 
frequency $\Omega$. Since we consider also heat-transport phenomena, we 
additionally include an infinitesimal heat current $Q$ as an external 
perturbation of the thermal equilibrium. Once the approach is set up, we 
can calculate several physical quantities explicitly. 

In Sec.\ \ref{S03} we calculate the entropy $S(T,\Omega)$ and the specific
heat $C_\Omega(T,\Omega)$ at constant rotation frequency $\Omega$. For nonzero
$\Omega$ we find that $S$ and $C_\Omega$ are smooth functions of the 
temperature $T$, so that the superfluid transition is a smooth crossover. 
From the maximum of the specific heat we obtain the transition temperature
$T_\lambda(\Omega)$ and calculate the shift $\Delta T_\lambda(\Omega)=
T_\lambda(\Omega)-T_\lambda$. We confirm the phase diagram of uniformly 
rotating liquid $^4$He shown in Fig.\ \ref{Fig01}. The correlation length 
$\xi=\xi(T,\Omega)$ is calculated in Sec.\ \ref{S04}. We show that $\xi$ is 
bounded by the mean distance between the vortex lines $L^{-1/2}$ and find 
a maximum at $T\approx T_\lambda(\Omega)$ as expected. 

In Sec.\ \ref{S05} we consider heat transport in uniformly rotating liquid
$^4$He and calculate the thermal-resistivity tensor $\rho_{\rm T}(T,\Omega)$. 
For $T\lesssim T_\lambda(\Omega)$ in the superfluid state the thermal 
resistivity is strongly anisotropic and depends on the direction of the heat 
current related to the rotation axis. In Sec.\ \ref{S06} we show that the 
thermal resistivity is caused by the mutual friction of the superfluid and the 
normal-fluid component due to the motion of the vortex lines, which was first
observed by Hall and Vinen \cite{HV}. We calculate the Vinen coefficients $B$
and $B^\prime$ and compare the results with a previous theory for the motion 
of vortex lines \cite{H2}. Finally, in Sec.\ \ref{S07} we compare the thermal
resistivity of uniformly rotating superfluid $^4$He with the thermal 
resistivity of nonrotating superfluid $^4$He in the presence of a nonzero
heat current $Q$, which was calculated previously in Ref.\ \onlinecite{H1}.
In this way we obtain the vortex density $L$ of the turbulent superfluid
flow induced by a finite heat current $Q$ which causes the Gorter-Mellink
mutual friction \cite{GM,V1}.

\section{Renormalization-group \break theory for rotating $^4$He} \label{S02}
\subsection{The model}
Dynamic critical phenomena in liquid $^4$He close to $T_\lambda$ are well 
described by model {\it F\,} which is given by the Langevin equations for the 
order parameter $\psi({\bf r},t)$ and the entropy variable $m({\bf r},t)$:
\begin{eqnarray}
{\partial \psi\over \partial t} &=& -2\Gamma_0 {\delta H\over \delta \psi^*}
+ ig_0 \psi {\delta H\over \delta m} + \theta_\psi \ ,
\label{B01} \\
{\partial m\over \partial t} &=& \lambda_0 \bbox{\nabla}^2 {\delta H\over
\delta m} - 2g_0 {\rm Im} \Bigl( \psi^* {\delta H\over \delta \psi^*} 
\Bigr) + \theta_m \ , \label{B02}
\end{eqnarray}
where
\begin{eqnarray}
H &=& \int d^d r \bigl[ {\textstyle\frac{1}{2}} \tau_0({\bf r}) 
\vert\psi\vert^2 + {\textstyle\frac{1}{2}} \vert (\bbox{\nabla}-i{\bf k})
\psi\vert^2 + \tilde u_0 \vert\psi\vert^4 \nonumber\\
&&\hskip1.1cm + {\textstyle\frac{1}{2}} \chi_0^{-1} m^2 + \gamma_0 m 
\vert\psi\vert^2 - h_0 m \bigr]  \label{B03} 
\end{eqnarray}
is the free energy functional and $\theta_\psi$ and $\theta_m$ are Gaussian
stochastic forces which incorporate the fluctuations. Because of the 
viscosity the normal-fluid component is rotating uniformly with velocity
${\bf v}_{\rm n} = {\bf \Omega}\times {\bf r}$. In (\ref{B03}) the uniform 
rotation is incorporated by the wave vector ${\bf k}$ which is related to 
the normal fluid velocity by 
\begin{equation}
{\bf k}= (m_4/\hbar)\, {\bf v}_{\rm n} = (m_4/\hbar)\,({\bf \Omega}\times 
{\bf r}) \ . 
\label{B04}
\end{equation}
We assume that the origin of the coordinate system is located on the rotation 
axis. The gravity and the centrifugal forces of the rotation imply a variation 
of the pressure in the helium according to 
\begin{equation}
P = P_0 - \rho [ gz - {\textstyle{1\over 2}} ({\bf \bf \Omega} \times 
{\bf r})^2 ] 
\label{B05}
\end{equation}
where $\rho$ is the density of liquid $^4$He and $g=981\ {\rm cm/s^2}$ is the
gravitational acceleration. Since the critical temperature $T_\lambda=
T_\lambda(P)$ depends on the pressure $P$, it depends on the space coordinate 
${\bf r}$ according to 
\begin{equation}
T_\lambda({\bf r}) = T_{\lambda 0} - (\partial T_\lambda / \partial P) \,
\rho [ gz - {\textstyle{1\over 2}} ({\bf \bf \Omega} \times {\bf r})^2 ] 
\label{B06}
\end{equation}
where \cite{A0} $T^\prime_\lambda = \rho g (-\partial T_\lambda /\partial P)
=+1.273\ {\rm \mu K/cm}$. In the energy functional (\ref{B03}) the spatially
dependent critical temperature (\ref{B06}) is represented by the parameter 
$\tau_0({\bf r})= 2\chi_0\gamma_0 [T_1-T_\lambda({\bf r})]$, where $T_1$ is 
an arbitrary constant reference temperature. 

Thus, the rotation influences the $^4$He in two different ways. First, in the
kinetic term of the energy functional (\ref{B03}) the wave vector (\ref{B04})
related to the normal fluid velocity generates straight vortex lines in the
superfluid component. Secondly, the centrifugal forces imply a slightly space
dependent critical temperature (\ref{B06}). In this paper we consider only
the first kind of influence which is first order in $\Omega$ and which is
the direct and most natural influence of the rotation. The second kind of
influence, which is indirect via the pressure variation and second order in
$\Omega$, will be neglected. Furthermore, we neglect gravity. In 
Sec.\ \ref{S03} we will show that these neglections are justified for 
realistic experiments performed in a microgravity environment in space. 
The experiments with rotating superfluid $^4$He to determine the Vinen 
coefficients \cite{HV,SL,L1,MS} were performed with a slow rotation rotation 
up to two turns per second. For this reason we assume that the rotation 
frequency is about $\Omega\approx 2\pi\ {\rm s^{-1}}$. 

For the calculation of the entropy and the specific heat in Sec.\ \ref{S03}
and of the correlation length in Sec.\ \ref{S04} we may consider the system 
in thermal equilibrium, where the temperature $T$ is constant. However, in 
Sec.\ \ref{S05} we consider heat transport and calculate the 
thermal-resistivity tensor in linear response. For this reason we must more 
generally consider the system in nonequilibrium, where an infinitesimal heat 
current ${\bf Q}$ is present which implies an infinitesimal temperature 
gradient $\bbox{\nabla}T$. Thus, the temperature $T({\bf r})$ depends weakly 
on the space coordinate ${\bf r}$. 

Model {\it F} is treated usually by field-theoretic means. The field-theoretic
perturbation theory and the renorma\-lization-group (RG) theory were developed 
by Dohm \cite{D1}. Our approach starts with the calculation of the
order-parameter Green's function in Hartree approximation and is described in 
detail in our previous paper (the second paper of Ref.\ \onlinecite{H1}).
While the approach was originally developed for liquid $^4$He in the presence
of a heat current $Q$, it is more generally valid and can be applied also
to rotating helium with a few modifications. The main task is the explicit
calculation of the Green's function, the set-up of the self-consistent
equations for the effective parameters, and the application of the RG theory. 

\subsection{Unrenormalized Green's function in Hartree approximation and 
self-consistent equations \break for the effective parameters}
In Hartree approximation the self energy includes only the self-consistent 
tadpole diagrams and hence is independent of momentum and frequency. 
Consequently, the Green's function has the same structure as the free Green's
function where only the parameters are replaced by the effective parameters.
For our calculations we need the equal times Green's function
\begin{equation}
G({\bf r},{\bf r}^\prime)= \langle\psi({\bf r},t)\psi^*({\bf r}^\prime,t)
\rangle  \ .
\label{B07}
\end{equation}
In Ref.\ \onlinecite{H1} we represented this Green's function in Hartree
approximation in terms of the integral
\begin{equation}
G({\bf r},{\bf r}^\prime) = 4\Gamma_0^\prime \int_0^\infty d\alpha 
\ e^{\alpha A} \, e^{\alpha B}  \,\delta({\bf r}-{\bf r}^\prime) 
\label{B08}
\end{equation}
where
\begin{eqnarray}
A&=&- \bigl\{ \Gamma_0 [r_1 - (\bbox{\nabla}-i{\bf k})^2] - 
i(g_0/2\chi_0\gamma_0) \Delta r_0 \bigr\}  \ ,  \label{B09}  \\
B&=&- \bigl\{ \Gamma_0^* [r_1 - (\bbox{\nabla}-i{\bf k})^2] + 
i(g_0/2\chi_0\gamma_0) \Delta r_0 \bigr\}  \label{B10}
\end{eqnarray}
are operators which are derived from the model-{\it F\,} equations 
(\ref{B01})-(\ref{B03}). Here ${\bf k}$ is the wave vector defined in 
(\ref{B04}) which incorporates the uniform rotation of the normal-fluid
component. $r_1$ and $\Delta r_0$ are the effective parameters. In 
Ref.\ \onlinecite{H1} it was shown that $r_1$ is related to the temperature 
parameter $r_0$ by 
\begin{equation}
r_1 = r_0 + 4 u_0 n_{\rm s} 
\label{B11}
\end{equation}
where
\begin{equation}
n_{\rm s} = \langle \vert\psi\vert^2 \rangle = G ({\bf r},{\bf r})  \ . 
\label{B12}
\end{equation}
The parameters $r_0$ and $\Delta r_0$ are related to the temperature 
$T({\bf r})$ and the critical temperature $T_\lambda$ by 
\begin{eqnarray}
r_0 &=& r_{\rm 0c} + 2\chi_0\gamma_0 \, [T({\bf r})-T_\lambda]/ T_\lambda \ ,
\label{B13}  \\
\Delta r_0 &=& 2\chi_0\gamma_0 \, [T({\bf r})-T_0]/ T_\lambda 
\label{B14}
\end{eqnarray}
where $r_{\rm 0c}=0$ in Hartree approximation and $T_0$ is an arbitrary 
constant reference temperature. (Compare (\ref{B11})-(\ref{B14}) with (3.31), 
(3.6), (3.30), and (3.26) in the second paper of Ref.\ \onlinecite{H1}, 
respectively). If $T({\bf r})$ and $T_\lambda$ are known, then $r_1$, $r_0$, 
and $\Delta r_0$ are determined by the self-consistent equations 
(\ref{B11})-(\ref{B14}). Since we neglect the pressure variations by 
gravity and by centrifugal forces, $T_\lambda$ is constant. For the 
calculations of the entropy, the specific heat, and the correlation length 
in Secs.\ \ref{S03} and \ref{S04} the system is assumed to be in thermal 
equilibrium so that the temperature $T({\bf r})=T$ is constant, too. Thus, 
in this case the effective parameters $r_1$, $r_0$, and $\Delta r_0$ are 
constant in space. 

For the calculation of the thermal-resistivity tensor in Sec.\ \ref{S05} the 
presence of an infinitesimal heat current ${\bf Q}$ is needed. Consequently,
there will be an infinitesimal temperature gradient $\bbox{\nabla}T$ which 
implies gradients of the effective parameters $\bbox{\nabla}r_1$, 
$\bbox{\nabla}r_0$, and $\bbox{\nabla}(\Delta r_0)$. Since $T_\lambda$ is 
assumed to be constant, Eqs.\ (\ref{B13}) and (\ref{B14}) imply
\begin{equation}
\bbox{\nabla}r_0 = \bbox{\nabla}(\Delta r_0) = 2\chi_0\gamma_0
\bbox{\nabla} T / T_\lambda \ . 
\label{B15}
\end{equation}
On the other hand, from (\ref{B11}) we obtain
\begin{equation}
\bbox{\nabla}r_1 = \bbox{\nabla}r_0 + 4u_0\bbox{\nabla} n_{\rm s} \ . 
\label{B16}
\end{equation}
Furthermore, we need a relation between the heat current and the gradients. 
For this purpose we take the average of the entropy equation (\ref{B02}) and 
obtain $\partial_t \langle m \rangle + \bbox{\nabla} {\bf q}=0$ where 
\begin{equation}
{\bf q} = - \lambda_0 \bbox{\nabla} \Bigl\langle {\delta H\over \delta m}
\Bigr\rangle - g_0 {\bf J}_{\rm s} 
\label{B17}
\end{equation}
is the entropy current which is related to the heat current ${\bf Q}$ in 
physical units by ${\bf q}={\bf Q}/k_{\rm B}T_\lambda$. The parameter
$\Delta r_0$ and the superfluid current ${\bf J}_{\rm s}$ are defined by
\cite{H1} 
\begin{eqnarray}
\Delta r_0 &=& 2\chi_0\gamma_0 \Bigl\langle {\delta H\over \delta m} 
\Bigr\rangle \ ,  \label{B18}  \\
{\bf J}_{\rm s} &=& \langle {\rm Im} [\psi^* (\bbox{\nabla}-i{\bf k}) \psi] 
\rangle =  \nonumber\\
&&\hskip-0.7cm = \bigl[ (2i)^{-1} [(\bbox{\nabla}-i{\bf k})-
(\bbox{\nabla}^\prime+i{\bf k}^\prime)] G({\bf r}, {\bf r}^\prime) 
\bigr]_{{\bf r}^\prime = {\bf r}} \ , 
\label{B19}
\end{eqnarray}
respectively. (Note that here the superfluid current is defined with respect
to the rotating frame.) Thus, from (\ref{B17}) we obtain 
\begin{equation}
{{\bf Q}\over g_0 k_{\rm B} T_\lambda}= {{\bf q}\over g_0}= - {\lambda_0\over 
g_0}\,{\bbox{\nabla}(\Delta r_0) \over 2\chi_0\gamma_0} - {\bf J}_{\rm s}  \ . 
\label{B20}
\end{equation}
If the heat current ${\bf Q}$ is given, then the gradients $\bbox{\nabla}
(\Delta r_0)$, $\bbox{\nabla}r_0$, $\bbox{\nabla}r_1$, and the temperature
gradient $\bbox{\nabla}T$ are determined by the self-consistent equations
(\ref{B15}), (\ref{B16}), and (\ref{B20}).

\subsection{Renormalization}
The quantities $n_{\rm s}$, $\bbox{\nabla} n_{\rm s}$, and ${\bf J}_{\rm s}$,
which appear in the first order terms of the self-consistent equations 
(\ref{B11}), (\ref{B16}), and (\ref{B20}), exhibit infrared divergencies 
at criticality where $r_1\rightarrow 0$ and $\Omega\rightarrow 0$. For
this reason, the self-consistent equations must be renormalized and the RG
theory must be applied to achieve a resummation of the infrared divergences
and a proper treatment of the critical fluctuations. We perform the
renormalization in the same way as in Ref.\ \onlinecite{H1} and use the
concept of renormalization by minimal subtraction of dimensional poles, which
is described for model {\it F\,} in Ref.\ \onlinecite{D1}. The calculations
are performed at fixed dimension $d=4-\epsilon$ (i.e.\ no $\epsilon$
expansion). For the renormalization of (\ref{B11}) we need the relations 
\cite{D1}
\begin{eqnarray}
r_0 - r_{\rm 0c} &=& Z_r r \ , 
\label{B21}  \\
u_0 &=& u\, Z_u Z_\psi^{-2} (\mu^\epsilon/A_d) \ . 
\label{B22}
\end{eqnarray}
In Hartree approximation it is $r_{\rm 0c}=0$ and the $Z$ factors are given
by \cite{H1}
\begin{equation}
Z_r = Z_u = 1/ [1 - 8u/\epsilon] 
\label{B23}
\end{equation}
and $Z_\psi=1$. Since $r_1$ is not renormalized \cite{H1}, from (\ref{B11})
we obtain 
\begin{equation}
r = r_1 [1 -8u/\epsilon] - 4u (\mu^\epsilon/A_d) n_{\rm s} \ . 
\label{B24}
\end{equation}
In the Appendix the Green's function $G({\bf r},{\bf r}^\prime)$ is 
evaluated for infinitesimal gradients of the effective parameters. Inserting
this Green's function into (\ref{B12}) we obtain
\begin{equation}
n_{\rm s}  = - {2\over\epsilon} A_d {\ell^{-2+\epsilon}\over 
\Gamma(-1+\epsilon/2)} \, {\cal F}_{-1+\epsilon/2}(r_1\ell^2) 
\label{B25}
\end{equation}
where ${\cal F}_p(\zeta)$ is defined by the integral 
\begin{equation}
{\cal F}_p(\zeta) = \int_0^\infty dv \, v^{p-1} {v\over \mbox{sh}\, v}
\, e^{-v\zeta}
\label{B26}
\end{equation}
and $\ell=(\hbar/2m_4\Omega)^{1/2}$ is the characteristic length related to
the mean distance between the vortices $L^{-1/2}$. (The vortex density is 
$L=1/2\pi\ell^2$.) Now, inserting (\ref{B25}) into 
(\ref{B24}) we obtain 
\begin{equation}
r \ell^2 = r_1 \ell^2 \bigl\{ 1 + 8u A \big\} 
\label{B27}
\end{equation}
where 
\begin{equation}
A = {1\over\epsilon} \Bigl[ {(\mu\ell)^\epsilon \over \Gamma(-1+\epsilon/2)}
(r_1\ell^2)^{-1} {\cal F}_{-1+\epsilon/2} (r_1\ell^2) - 1 \Bigr] \ . 
\label{B28}
\end{equation}
Eq.\ (\ref{B27}) is the renormalized counterpart of (\ref{B11}). We have 
multiplied both sides by $\ell^2$ because $r\ell^2$ and $r_1\ell^2$ are
dimensionless quantities.

Next, by applying $\bbox{\nabla}$ to (\ref{B27}) we obtain the renormalized
counterpart of (\ref{B16}). We find
\begin{equation}
\bbox{\nabla}r\ell^3 = (\bbox{\nabla}r_1\ell^3) \bigl\{ 1 + 8u A_1 \bigr\} 
\label{B29}
\end{equation}
where
\begin{equation}
A_1 = {1\over\epsilon} \Bigl[{-(\mu\ell)^\epsilon \over \Gamma(-1+\epsilon/2)}
{\cal F}_{\epsilon/2} (r_1\ell^2) - 1 \Bigr] \ . 
\label{B30}
\end{equation}
The factors $\ell^3$ are multiplied on both sides of (\ref{B29}) because 
$\bbox{\nabla}r\ell^3$ and $\bbox{\nabla}r_1\ell^3$ are dimensionless 
quantities.

For the renormalization of (\ref{B20}) we need the rela\-tions \cite{D1} 
\begin{eqnarray}
\chi_0\gamma_0 &=& \gamma\, (\chi_0 Z_m)^{1/2}Z_r (\mu^\epsilon/A_d)^{1/2} \ , 
\label{B31}  \\
g_0 &=& g\, (\chi_0 Z_m)^{1/2} (\mu^\epsilon/A_d)^{1/2} \ ,  \label{B32}  \\
\lambda_0 &=& \chi_0 Z_\lambda^{-1} \lambda \ ,  \label{B33}
\end{eqnarray}
and furthermore $\Delta r_0= Z_r \Delta r$. The entropy current is 
renormalized by ${\bf q}=(\chi_0 Z_m)^{1/2} {\bf q}_{\rm ren}$. Consequently,
the $Z$ factors cancel in the ratio ${\bf q}/g_0= ({\bf q}_{\rm ren}/g)
(\mu^\epsilon/A_d)^{1/2}$, so that the left-hand side of (\ref{B20}) needs not
be renormalized. On the right-hand side of (\ref{B20}) part of the $Z$ 
factors cancel. Thus, we obtain
\begin{equation}
{{\bf Q} \over g_0 k_{\rm B} T_\lambda} = - {A_d\over \mu^\epsilon}\,
{\lambda\over g}\, {\bbox{\nabla}(\Delta r)\over 2\gamma}\, 
(Z_m Z_\lambda)^{-1}  -  {\bf J}_{\rm s} \ . 
\label{B34}
\end{equation}
The dynamic couplings of model {\it F\,} like $\lambda$ and $g$ occur only in 
dimensionless combinations like $\lambda/g$. For this reason, we express 
(\ref{B34}) entirely in terms of the dimensionless couplings \cite{D1}
$w=\Gamma/\lambda$, $F=g/\lambda$, and $f=g^2/\lambda\Gamma^\prime=F^2/
w^\prime$. In Hartree approximation the $Z$ factors are given by \cite{H1}
\begin{equation}
Z_m Z_\lambda = 1/[1-f/2\epsilon]  \ . 
\label{B35}
\end{equation}
Thus, we rewrite (\ref{B34}) in the form
\begin{equation}
{{\bf Q}/\mu^{d-1} \over g_0 k_{\rm B} T_\lambda} = - {A_d\over 2\gamma F}\,
{\bbox{\nabla}(\Delta r)\over \mu^3}\, \Bigl[ 1- {f\over 2\epsilon} \Bigr]
- {{\bf J}_{\rm s} \over \mu^{d-1}} \ . 
\label{B36}
\end{equation}
The superfluid current ${\bf J}_{\rm s}$ is obtained from (\ref{B19}) by 
inserting the Green's function $G({\bf r},{\bf r}^\prime)$, which we have 
calculated explicitly in the Appendix. We find
\begin{eqnarray}
{\bf J}_{\rm s} &=& {1\over \epsilon} A_d {\ell^\epsilon\over \Gamma(-1+
\epsilon/2)} \, \Bigl\{ {\cal M}_{\epsilon/2}(r_1\ell^2)\, [{\bf e}_z \times
\bbox{\nabla}r_1] \nonumber\\
&&- {F\over 4\gamma w^\prime} \Bigl[ {\cal F}_{\epsilon/2}(r_1\ell^2) 
\, {\bf e}_z ({\bf e}_z \cdot \bbox{\nabla}(\Delta r))  \nonumber\\
&&\hskip1cm - {\cal N}^\prime_{\epsilon/2}(r_1\ell^2) \,[{\bf e}_z\times 
[{\bf e}_z\times \bbox{\nabla}(\Delta r)]]  \nonumber\\
&&\hskip1cm + {\cal N}^{\prime\prime}_{\epsilon/2}(r_1\ell^2) 
\,[{\bf e}_z\times \bbox{\nabla}(\Delta r)]  \Bigr] \Bigr\} 
\label{B37}
\end{eqnarray}
where ${\cal F}_p(\zeta)$ is defined in (\ref{B26}) and ${\cal M}_p(\zeta)$, 
${\cal N}^\prime_p(\zeta)$, and ${\cal N}^{\prime\prime}_p(\zeta)$ are given 
by the integrals 
\begin{eqnarray}
{\cal M}_p(\zeta) &=& \int_0^\infty dv\, {v^{p-1}\over \mbox{sh}\, v}
\Bigl( {v\over\mbox{th}\, v} -1 \Bigr) \, e^{-v\zeta} \ ,  \label{B38} \\
{\cal N}^\prime_p(\zeta) &=& {2\over 1+(w^{\prime\prime}/w^\prime)^2}
\int_0^\infty dv\, {v^{p-1}\over \mbox{sh}^2 v} \nonumber\\
&&\hskip0.8cm \times\Bigl[ \mbox{ch}\, v- \cos\Bigl({w^{\prime\prime}\over 
w^\prime}v\Bigr)\Bigr] \, e^{-v\zeta} \ ,  \label{B39} \\
{\cal N}^{\prime\prime}_p(\zeta) &=& -{2\over 1+(w^{\prime\prime}/w^\prime)^2}
\int_0^\infty dv\, {v^{p-1}\over \mbox{sh}^2 v} \nonumber\\
&&\hskip0.8cm \times\Bigl[ {w^{\prime\prime}\over w^\prime} \mbox{sh}\, v- 
\sin\Bigl({w^{\prime\prime}\over w^\prime}v\Bigr)\Bigr] \, e^{-v\zeta} \ , 
\label{B40} 
\end{eqnarray}
respectively. Now, inserting (\ref{B37}) into (\ref{B36}) and replacing
$\bbox{\nabla}r_1$ in terms of $\bbox{\nabla}(\Delta r)$ by using (\ref{B29})
and $\bbox{\nabla}r= \bbox{\nabla}(\Delta r)$, we eventually obtain
\begin{eqnarray}
{{\bf Q}/\mu^{d-1} \over g_0 k_{\rm B} T_\lambda} &=& - {A_d\over 2\gamma F}\,
\Bigl\{ \Bigl[ 1 + {f\over 2} A_1\Bigr] \,{\bf e}_z ({\bf e}_z\cdot 
\bbox{\nabla}(\Delta r))/ \mu^3   \nonumber\\
&&\hskip1.3cm - \Bigl[ 1 + {f\over 2} A^\prime\Bigr] [{\bf e}_z\times
[{\bf e}_z\times \bbox{\nabla}(\Delta r)]]/ \mu^3  \nonumber\\
&&\hskip1.3cm + {f\over 2} A^{\prime\prime} [{\bf e}_z\times \bbox{\nabla}
(\Delta r)] /\mu^3 \Bigr\}
\label{B41}
\end{eqnarray}
where $A_1$ is defined in (\ref{B30}) and 
\begin{eqnarray}
A^\prime &=& {1\over \epsilon} \Bigr[ {-(\mu\ell)^\epsilon\over \Gamma(-1+ 
\epsilon/2)} \, {\cal N}^\prime_{\epsilon/2}(r_1\ell^2) -1 \Bigr] \ , 
\label{B42} \\
A^{\prime\prime} &=& {1\over \epsilon}\, {-(\mu\ell)^\epsilon\over \Gamma(-1+ 
\epsilon/2)} \, \Bigr[ {\cal N}^{\prime\prime}_{\epsilon/2}(r_1\ell^2)
\nonumber\\
&&\hskip2.5cm - {4\gamma w^\prime/F \over 1 + 8u A_1} {\cal M}_{\epsilon/2}
(r_1\ell^2) \Bigr] \ .  \label{B43}
\end{eqnarray}
Finally, we renormalize (\ref{B13})-(\ref{B15}). Inserting (\ref{B21}) and
(\ref{B31}) into (\ref{B13}) we obtain 
\begin{equation}
r = 2\gamma (\chi_0 Z_m)^{1/2} (\mu^\epsilon/A_d)^{1/2} [T({\bf r})-T_\lambda]
/T_\lambda \ . 
\label{B44}
\end{equation}
Clearly, in this equation the $Z$ factor $(\chi_0 Z_m)^{1/2}$ does not cancel.
For convenience we define the parameter
\begin{equation}
\tau = \Bigl( {A_d \mu^d \over\chi_0 Z_m} \Bigr)^{1/2} \,{1\over 2\gamma} \ , 
\label{B45}
\end{equation}
so that (\ref{B44}) is written in the simple form
\begin{equation}
r/\mu^2 = \tau^{-1} [T({\bf r}) - T_\lambda] / T_\lambda \ . 
\label{B46}
\end{equation}
The renormalized counterparts of (\ref{B14}) and (\ref{B15}) are obtained
analogously. We find
\begin{equation}
\Delta r/\mu^2 = \tau^{-1} [T({\bf r})-T_0] / T_\lambda \ , 
\label{B47} 
\end{equation}
\vspace*{-0.7cm}
\begin{equation}
\bbox{\nabla} r/\mu^3 = \bbox{\nabla} (\Delta r)/\mu^3 = 
\tau^{-1} (\mu T_\lambda)^{-1} \bbox{\nabla}T \ , 
\label{B48} 
\end{equation}
respectively.

\subsection{Application of the RG theory}
By the renormalization a characteristic length scale is introduced which is
described by the parameter $\mu$. The RG theory is based on the fact that 
this length scale is arbitrary and may be changed according to 
$\mu\rightarrow \mu l$, where $l$ is the RG flow parameter. As a consequence, 
the renormalized coupling parameters $u(l)$, $\gamma(l)$, $w(l)$, $F(l)$, and 
$f(l)$ depend on $l$. Furthermore, also the $Z$ factors depend on $l$. 
(Note that the RG flow parameter $l$ must be distinguished from the 
characteristic length $\ell=(\hbar/2m_4\Omega)^{1/2}$.) Now, the dimensionless 
parameter defined in (\ref{B45}) reads
\begin{equation}
\tau = \Bigl( {A_d (\mu l)^d \over\chi_0 Z_m(l)} \Bigr)^{1/2} 
\,{1\over 2\gamma(l)} \ . 
\label{B49}
\end{equation}
For convenience we will use $\tau$ as the RG flow parameter instead of 
$l$ because $\tau$ is closely related to the reduced temperature by 
(\ref{B46}) and the renormalized coupling parameters $u[\tau]$, 
$\gamma[\tau]$, $w[\tau]$, $F[\tau]$, and $f[\tau]$ were determined as
functions of $\tau$ in Ref.\ \onlinecite{D1}. We identify $\mu l = \xi^{-1}$
by the correlation length $\xi=\xi(\tau)$, which in the asymptotic region
is given by $\xi(\tau)=\xi_0 \tau^{-\nu}$. The identification $\mu l = 
\xi^{-1}$ is correct in one-loop order, corrections appear in higher orders 
\cite{SD}.

Now, we write the self-consistent equations for the effective parameters in
a form which is appropriate for the numerical evaluation. From (\ref{B27}),
(\ref{B29}), and (\ref{B41}) we obtain
\begin{eqnarray}
r \ell^2 &=& r_1 \ell^2 \bigl\{ 1 + 8u[\tau] A \big\}  \ ,  \label{B50}  \\
\bbox{\nabla}r\ell^3 &=& (\bbox{\nabla}r_1\ell^3) \bigl\{ 1 + 8u[\tau] A_1 
\bigr\}  \ ,  \label{B51}  \\
{{\bf Q}\, \xi^{d-1} \over g_0 k_{\rm B} T_\lambda} &=& - {A_d\over 
2\gamma[\tau] F[\tau]}\, \bigl\{ [ 1 + {\textstyle{1\over 2}} f[\tau] 
A_1] \,{\bf e}_z ({\bf e}_z\cdot \bbox{\nabla}(\Delta r)) \, \xi^3  
\nonumber\\
&&\hskip1.1cm - [ 1 + {\textstyle{1\over 2}} f[\tau] A^\prime] 
[{\bf e}_z\times [{\bf e}_z\times \bbox{\nabla}(\Delta r)]] \, \xi^3  
\nonumber\\
&&\hskip1.1cm + {\textstyle{1\over 2}} f[\tau] A^{\prime\prime} 
[{\bf e}_z\times \bbox{\nabla} (\Delta r)] \, \xi^3 \bigr\}  \ .  \label{B52}
\end{eqnarray}
We replace $\mu\ell \rightarrow \mu l\,\ell= \ell/\xi$ in (\ref{B28}), 
(\ref{B30}), (\ref{B42}), and (\ref{B43}) and obtain the amplitudes 
\begin{eqnarray}
A &=& {1\over\epsilon}\Bigl[ {(\ell/\xi)^\epsilon \over \Gamma(-1+\epsilon/2)}
(r_1\ell^2)^{-1} {\cal F}_{-1+\epsilon/2} (r_1\ell^2) - 1 \Bigr] \ ,  
\hskip-0.2cm  \label{B53}  \\
A_1 &=& {1\over\epsilon} \Bigl[{-(\ell/\xi)^\epsilon \over \Gamma(-1+
\epsilon/2)} {\cal F}_{\epsilon/2} (r_1\ell^2) - 1 \Bigr] \ ,  \label{B54} \\
A^\prime &=& {1\over \epsilon} \Bigr[ {-(\ell/\xi)^\epsilon\over \Gamma(-1+ 
\epsilon/2)} \, {\cal N}^\prime_{\epsilon/2}(r_1\ell^2) -1 \Bigr] \ , 
\label{B55} \\
A^{\prime\prime} &=& {1\over \epsilon}\, {-(\ell/\xi)^\epsilon\over \Gamma(-1+ 
\epsilon/2)} \, \Bigr[ {\cal N}^{\prime\prime}_{\epsilon/2}(r_1\ell^2)
\nonumber\\
&&\hskip1.7cm - {4\gamma[\tau] w^\prime[\tau]/F[\tau] \over 1 + 8u[\tau] A_1} 
{\cal M}_{\epsilon/2} (r_1\ell^2) \Bigr] \ .  \label{B56}
\end{eqnarray}
The functions ${\cal F}_p(\zeta)$, ${\cal M}_p(\zeta)$, ${\cal N}^\prime_p
(\zeta)$, and ${\cal N}^{\prime\prime}_p(\zeta)$ are defined by the integrals
(\ref{B26}) and (\ref{B38})-(\ref{B40}) as before, where however in 
(\ref{B39}) and (\ref{B40}) the parameters $w^\prime$ and $w^{\prime\prime}$
must be replaced by $w^\prime[\tau]$ and $w^{\prime\prime}[\tau]$. Finally,
from (\ref{B46})-(\ref{B48}) we obtain the equations 
\begin{eqnarray}
r\, \xi^2 &=& \tau^{-1} [T({\bf r}) - T_\lambda] / T_\lambda \ , 
\label{B57}  \\
\Delta r\,\xi^2 &=& \tau^{-1} [T({\bf r})-T_0]/ T_\lambda \ , \label{B58}  \\
\bbox{\nabla} r\, \xi^3 &=& \bbox{\nabla} (\Delta r) \,  \xi^3 = 
\tau^{-1} (\xi/ T_\lambda) \bbox{\nabla}T 
\label{B59} 
\end{eqnarray}
which relate the effective parameters to the temperature $T({\bf r})$ and 
its gradient.

Supposed the temperature $T$, the heat current ${\bf Q}$, and the rotation
frequency $\Omega$ are known quantities, then Eqs.\ (\ref{B50})-(\ref{B59})
are ten equations for the eleven unknown variables $r_1$, $r$, $\Delta r$,
$\bbox{\nabla}r_1$, $\bbox{\nabla}r$, $\bbox{\nabla}(\Delta r)$, $A$, $A_1$,
$A^\prime$, $A^{\prime\prime}$, and $\tau$. Thus, one variable remains 
undetermined which actually is the RG flow parameter $\tau$. In the spirit of
the RG theory we must choose the flow parameter $\tau$ so that in the 
perturbation series the infrared divergencies are resummed in an optimum way. 
By experience we find that the condition
\begin{eqnarray}
&&[r_1\ell^2 + 1] (\xi/\ell)^2 \nonumber\\ 
&&+ (32u[\tau]/\epsilon)(1-\epsilon/2)
(\xi/\ell)^{2-\epsilon} [r_1\ell^2 + 1]^{-\epsilon/2} = 1 
\label{B60}
\end{eqnarray}
is a good choice for fixing $\tau$. 

The integrals (\ref{B26}) and (\ref{B38})-(\ref{B40}) are well defined for
$\zeta=r_1\ell^2$ in the interval $-1<\zeta<+\infty$. For $\zeta\gg+1$ and
for $\zeta\rightarrow -1$ the integrals can be evaluated asymptotically.
From (\ref{B26}) we obtain
\begin{equation}
{\cal F}_p(\zeta) \approx \cases{ \Gamma(p)\, \zeta^{-p}  &for $\zeta\gg +1$\, 
, \cr  2 \Gamma(p+1) (\zeta+1)^{-(p+1)}  &for $\zeta\rightarrow -1$\, . \cr }
\label{B61}
\end{equation}
Consequently, from (\ref{B50}) together with (\ref{B53}) we obtain
\begin{equation}
r\xi^2 = \cases{ r_1\xi^2 \{ 1 + (8u[\tau]/\epsilon)
[ (r_1\xi^2)^{-\epsilon/2} - 1 ] \}  \hspace*{-2.5cm} \cr 
&for $r_1\ell^2 \gg +1$  , \vspace*{0.1cm}  \cr
-(16u[\tau]/\epsilon) (1-\epsilon/2) (\xi/\ell)^{2-\epsilon}
[r_1\ell^2 +1 ]^{-\epsilon/2}  \hspace*{-2.5cm} \cr 
&for $r_1\ell^2 \rightarrow -1$ . \cr }
\label{B62}
\end{equation}
For temperatures $T$ well above $T_\lambda$ the correlation length $\xi$
decreases so that $\xi\ll\ell$. Eq.\ (\ref{B60}) implies $r_1\ell^2\gg +1$
so that on the left-hand side of (\ref{B60}) the first term dominates while 
the second term can be neglected. Thus, Eq.\ (\ref{B60}) reduces into 
$r_1\xi^2=1$ and Eq.\ (\ref{B62}) implies $r\ell^2=r_1\ell^2$. Consequently,
for $T\gg T_\lambda$ the flow-parameter condition (\ref{B60}) reduces into
\begin{equation}
r\xi^2 = 1 \ . 
\label{B63}
\end{equation}
On the other hand, for temperatures $T$ well below $T_\lambda$ it is 
$r_1\ell^2\rightarrow -1$, so that on the left-hand side of (\ref{B60}) now 
the second term dominates while the first term is small and negligible. 
Again it is $\xi\ll\ell$. Consequently, (\ref{B62}) implies that for 
$T\ll T_\lambda$ the flow parameter condition (\ref{B60}) reduces into 
\begin{equation}
-2 r\xi^2 = 1 \ . 
\label{B64}
\end{equation}
Actually, (\ref{B63}) and (\ref{B64}) are the standard flow-para\-meter 
conditions of Ref.\ \onlinecite{D1} for $T>T_\lambda$ and $T<T_\lambda$, 
respectively, where $Q=0$ and $\Omega=0$. (Note that our $r\xi^2$ is 
identified by $r(l)/(\mu l)^2$ in Ref.\ \onlinecite{D1}.) Thus, we have shown 
that our flow parameter condition (\ref{B60}), which is valid for finite 
$\Omega$ and infinitesimal small $Q$, reduces to the standard flow parameter 
conditions of Ref.\ \onlinecite{D1} for temperatures $T$ well above and well 
below $T_\lambda$. For $T$ near $T_\lambda$ it represents an interpolation.

Now, Eqs.\ (\ref{B50})-(\ref{B60}) are eleven equations which determine the
eleven unknown variables uniquely. We solve these equations numerically for
given temperature $T$, heat current $Q$, and rotation frequency $\Omega$ to
determine the effective parameters. From (\ref{B59}) we obtain the temperature
gradient $\bbox{\nabla}T$. Once the effective parameters are known, physical
quantities can be calculated explicitly. This will be done in the next 
section. Since we consider liquid $^4$He in three dimensions, we set $d=3$
and $\epsilon=4-d=1$ in all formulas when performing the numerical 
calculations.

As an input we need the dimensionless renormalized couplings $u[\tau]$,
$\gamma[\tau]$, $w[\tau]=w^\prime[\tau] + i w^{\prime\prime}[\tau]$, 
$F[\tau]$, and $f[\tau]$ as functions of $\tau$ which have been determined 
by Dohm \cite{D1}. Furthermore, we need the parameter $g_0$ which is related 
to the entropy at $T_\lambda$. For liquid helium at saturated vapor pressure
this parameter is \cite{TA} $g_0=2.164\times 10^{11}\ {\rm s}^{-1}$. To
calculate the correlation length $\xi(\tau)=\xi_0 \tau^{-\nu}$ as a function
of $\tau$ we use the exponent $\nu=0.671$ and the amplitude $\xi_0=1.45\times
10^{-8}\ {\rm cm}$ which were determined experimentally in 
Refs.\ \onlinecite{TA} and \onlinecite{LS}. There are no adjustable 
parameters.

\section{Entropy and specific heat} \label{S03}
In our previous paper \cite{H1} the entropy was considered in Hartree
approximation combined with the renorma\-lization-group theory. The following 
result was obtained:
\begin{equation}
S = S_\lambda + t\,\bigl\{ B + \tilde A\,[\,(4\nu / \alpha) + E[u^*]\,]
\tau^{-\alpha} \bigr\} 
\label{C01}
\end{equation}
where
\begin{eqnarray}
t&=& \tau\,r\xi^2 = [T-T_\lambda]/T_\lambda \ , \label{C02} \\
E[u] &=& (2u)^{-1} [1-\rho_1/\rho] \ . \label{C03} 
\end{eqnarray}
This formula is valid also for rotating $^4$He without alteration. We just 
insert the effective parameters $r$ and $r_1$ and the RG flow parameter $\tau$
determined in Sec.\ \ref{S02}. Since the entropy is an equilibrium quantity, 
the infinitesimal gradients may be discarded so that only the self-consistent
equations (\ref{B50}), (\ref{B53}), and (\ref{B57}) must be solved. The 
specific heat $C_\Omega$ is then obtained by numerical differentiation with
respect to the temperature according to
\begin{equation}
C_\Omega = T_\lambda \, \Bigl( {\partial S \over \partial T} \Bigr)_\Omega
= \Bigl( {\partial S \over \partial t} \Bigr)_\Omega 
\label{C04}
\end{equation}
where the rotation frequency ${\bf \Omega}=\Omega {\bf e}_z$ is kept constant.

The formula (\ref{C01}) is derived in the asymptotic regime close to 
$T_\lambda$ where $\xi=\xi_0 \tau^{-\nu}$ and $u[\tau]\approx u^*=0.0362$. 
The critical exponents $\nu=0.671$ and $\alpha=-0.013$ and the nonuniversal
amplitudes $\tilde A=2.22\ {\rm J/mol\, K}$ and $B=456\ {\rm J/mol\, K}$ are 
obtained by comparing the theoretical specific heat for $\Omega=0$ defined by 
(\ref{C01})-(\ref{C04}) with the most recent experimental data \cite{LS},
which were obtained in a microgravity environment in space.

\begin{figure}[t]
\vspace*{6.9cm}
\includegraphics{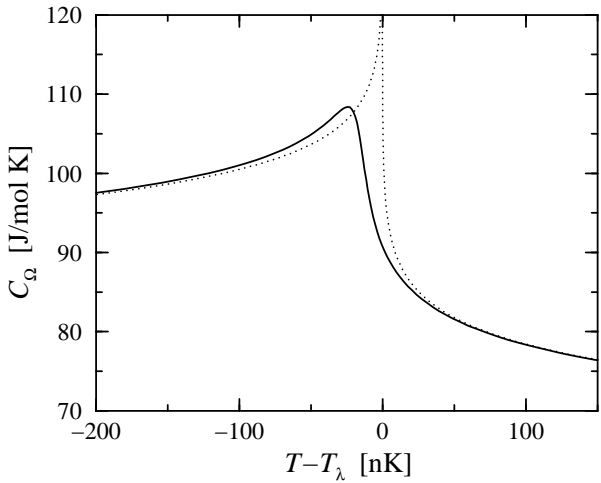}
\caption{The specific heat for rotating $^4$He as a function of temperature. 
The solid line represents our theoretical result for $\Omega=2\pi\ {\rm 
s^{-1}}$. For comparison, the specific heat for $\Omega=0$ is shown as dotted 
line.}
\label{Fig02}
\end{figure}

Realistic experiments with rotating $^4$He were performed for rotations about
one turn per second (see Refs.\ \onlinecite{HV,SL,L1,MS}). For this reason, 
we calculate the specific heat for $\Omega=2\pi\ {\rm s^{-1}}$. In 
Fig.\ \ref{Fig02} our numerical result is shown as solid line. For comparison 
the dotted line represents the specific heat at $\Omega=0$. Clearly, for 
nonzero $\Omega$ the specific heat is a smooth function of temperature with a 
maximum slightly below $T_\lambda$. The position of the maximum can be 
interpreted as the superfluid transition temperature $T_\lambda(\Omega)$ for 
rotating $^4$He which, however, is not sharply defined. Thus, for nonzero
$\Omega$ the superfluid transition is a smooth crossover. The shift of the
transition temperature $\Delta T_\lambda(\Omega)=T_\lambda(\Omega)-T_\lambda$
is negative. For $\Omega=2\pi\ {\rm s^{-1}}$ we find $\Delta T_\lambda(\Omega)
= -25\ {\rm nK}$ from Fig.\ \ref{Fig02}. For other rotation frequencies 
$\Omega$ a scaling formula for $\Delta T_\lambda(\Omega)$ can be derived. 
Scaling requires the relations $\xi/\ell=a$ and $\Delta T_\lambda(\Omega)/
T_\lambda=b\,\tau$ where $a$ and $b$ are dimensionless constants. Inserting
$\xi=\xi_0 \tau^{-\nu}$ and $\ell=(\hbar/2m_4\Omega)^{1/2}$, eliminating 
$\tau$, and solving for $\Delta T_\lambda(\Omega)$ we obtain
\begin{equation}
\Delta T_\lambda(\Omega) = - M_\lambda T_\lambda (2m_4\xi_0^2 \Omega/\hbar
)^{1/2\nu}
\label{C05}
\end{equation}
where $M_\lambda$ is a dimensionless constant of order unity related to $a$ 
and $b$. From the position of the maximum in Fig.\ \ref{Fig02} we find 
$M_\lambda=1.2$. The shift of the critical temperature (\ref{C05}) is 
represented in Fig.\ \ref{Fig01} by the dashed line. Thus, the phase diagram 
for uniformly rotating liquid $^4$He shown in Fig.\ \ref{Fig01} is confirmed.

Our theory neglects the influence of the centrifugal forces which imply
a spatial variation of the critical temperature given by (\ref{B06}). For
a sample of diameter $D$ the maximum variation of the critical temperature is 
\begin{equation}
\delta T_{\lambda{\rm ,max}}(\Omega) = \vert \partial T_\lambda / \partial P 
\vert \, \rho \, \Omega^2 D^2/8 \ . 
\label{C06}
\end{equation}
To justify the neglection of the influence of the centrifugal forces we must
require that $\delta T_{\lambda{\rm ,max}}(\Omega)$ is ten times smaller
than $\Delta T_\lambda(\Omega)$ defined in (\ref{C05}). Thus, for $\Omega= 
2\pi\ {\rm s^{-1}}$ we require $\delta T_{\lambda{\rm ,max}}(\Omega)=2.5
\ {\rm nK}$. Inserting this value into (\ref{C06}) and solving for $D$ we 
obtain the maximum diameter $D_{\rm max}$ of the sample for which the 
influences of the centrifugal forces may be neglected. For $\Omega=2\pi
\ {\rm s^{-1}}$ we obtain $D_{\rm max}=0.6\ {\rm cm}$, while from (\ref{C05}) 
and (\ref{C06}) we find the frequency dependence $D_{\rm max}\sim 
\Omega^{1/4\nu-1}=\Omega^{-0.627}$. Thus, $D_{\rm max}$ is of order of 
realistic sample sizes. If the samples are made smaller so that $D\lesssim 
D_{\rm max}$ is satisfied, then indeed the influence of the centrifugal 
forces may be neglected. 

On the other hand the influences of gravity on earth imply a much larger
variation $\delta T_{\lambda g}$ of the critical temperature. From (\ref{B06})
we obtain
\begin{equation}
\delta T_{\lambda g} = \vert \partial T_\lambda / \partial P \vert \,
\rho g\, \Delta z 
\label{C07}
\end{equation}
where $\Delta z$ is the height of the sample. For $\Delta z\approx 0.6
\ {\rm cm}$ we obtain $\delta T_{\lambda g}\approx 0.8\ {\rm \mu K}$ which
is much larger than the temperature scale in Fig.\ \ref{Fig02}. For this
reason, the maximum of the specific heat $C_\Omega$ induced by the vortices
cannot be resolved in an experiment on earth. The experiment must be 
performed in a microgravity environment in space.

For temperatures well below $T_\lambda$ a first-order transition between a 
vortex liquid and a vortex lattice is expected, because for decreasing 
temperatures the fluctuations decrease. However, our theory based on the 
Hartree approximation cannot describe this transition. While the effects of 
vortices are included indirectly, our theory assumes the physical quantities 
to be homogeneous in space so that the system is always a vortex liquid.

\section{Correlation length} \label{S04}
\begin{figure}[t]
\vspace*{6.9cm}
\includegraphics{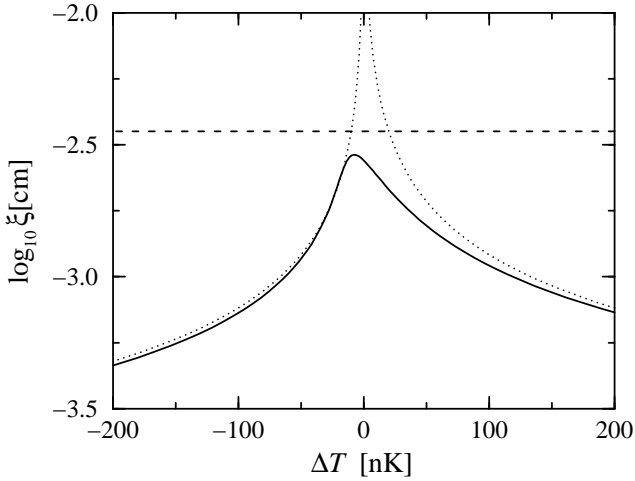}
\caption{The correlation length for rotating $^4$He as a function of 
temperature. The solid line is the correlation length $\xi$ for $\Omega=2\pi
\ {\rm s^{-1}}$. The dashed line represents the magnetic length $\ell$. For 
comparison, the correlation length $\xi$ for $\Omega=0$ is shown as dotted 
line.}
\label{Fig03}
\end{figure}

In the asymptotic regime close to $T_\lambda(\Omega)$ the correlation length 
is given by $\xi=(\mu l)^{-1}=\xi_0 \tau^{-\nu}$. Inserting the RG flow 
parameter $\tau$ determined in Sec.\ \ref{S02}, we calculate 
$\xi=\xi(T,\Omega)$ as a function of temperature $T$ for the rotation 
frequency $\Omega=2\pi\ {\rm s}^{-1}$. Our numerical result is shown in 
Fig.\ \ref{Fig03} as solid line. For comparison, the dotted line represents 
the correlation length at $\Omega=0$. Clearly, for nonzero $\Omega$ the 
correlation length is a smooth function of temperature with a maximum 
slightly below $T_\lambda$. This maximum is located close to the maximum of 
the specific heat in Fig.\ \ref{Fig02}.

On the other hand, the characteristic length $\ell=(\hbar/2m_4\Omega)^{1/2}$
is shown in Fig.\ \ref{Fig03} as dashed line. By definition this length does
not depend on the temperature. We find $\xi<\ell$ so that the correlation
length is bounded by $\ell$. Near the maximum it is $\xi\approx\ell$.
The characteristic length $\ell$ is closely related to the mean distance
between the vortex lines $L^{-1/2}$. Because of $L=1/2\pi\ell^2$ it is 
$L^{-1/2}=(2\pi)^{1/2} \ell\approx 2.5\,\ell$. Consequently, the correlation
length $\xi$ is bounded by the mean distance between the vortex lines 
according to $\xi<L^{-1/2}$.

\section{Thermal conductivity \break and resistivity} \label{S05}
We eliminate the gradient $\bbox{\nabla}(\Delta r)$ from (\ref{B52}) and 
(\ref{B59}) and solve the resulting equation for ${\bf Q}$.
Then we obtain
\begin{eqnarray}
{\bf Q} &=& - \lambda_{\rm T,1} {\bf e}_z ({\bf e}_z\cdot \bbox{\nabla}T) 
\vspace*{0.1cm}  \nonumber\\
&&+ \lambda^\prime_{\rm T} ({\bf e}_z\times ({\bf e}_z\times \bbox{\nabla}T))
 - \lambda^{\prime\prime}_{\rm T} ({\bf e}_z \times \bbox{\nabla}T)
\label{D01}
\end{eqnarray}
where
\begin{eqnarray}
\lambda_{\rm T,1} &=& {g_0 k_{\rm B} A_d \over \tau \, \xi^{d-2} } \, 
{ \{ 1 + (f[\tau]/2) A_1 \} \over 2 \gamma[\tau] F[\tau] }  \ , 
\label{D02}  \\
\lambda^\prime_{\rm T} &=& {g_0 k_{\rm B} A_d \over \tau \, \xi^{d-2} } \, 
{ \{ 1 + (f[\tau]/2) A^\prime \} \over 2 \gamma[\tau] F[\tau] }  \ , 
\label{D03}  \\
\lambda^{\prime\prime}_{\rm T} &=& {g_0 k_{\rm B} A_d\over \tau \, \xi^{d-2} } 
\, { (f[\tau]/2) A^{\prime\prime} \over 2 \gamma[\tau] F[\tau] }  
\label{D04}
\end{eqnarray}
are three components of the thermal-conductivity tensor. While the dependence
on the dimensionality $d$ is needed for the renormalization, the formulas
(\ref{D02})-(\ref{D04}) are evaluated for $d=3$. Clearly, Eq.\ (\ref{D01}) is 
a linear relation between the temperature gradient $\bbox{\nabla}T$ and the 
heat current ${\bf Q}$. Solving this equation for $\bbox{\nabla}T$ we obtain
\begin{eqnarray}
\bbox{\nabla}T &=& - \rho_{\rm T,1} {\bf e}_z ({\bf e}_z\cdot {\bf Q}) 
\vspace*{0.1cm}  \nonumber\\
&&+ \rho^\prime_{\rm T} ({\bf e}_z\times ({\bf e}_z\times {\bf Q}))
 - \rho^{\prime\prime}_{\rm T} ({\bf e}_z \times {\bf Q})
\label{D05}
\end{eqnarray}
where $\rho_{\rm T,1}$, $\rho^\prime_{\rm T}$, and 
$\rho^{\prime\prime}_{\rm T}$ are three components of the thermal-resistivity
tensor which are related to the conductivities (\ref{D02})-(\ref{D04}) by 
\begin{eqnarray}
\rho_{\rm T,1} &=& 1/ \lambda_{\rm T,1} \ ,  \label{D06}  \\
\rho^\prime_{\rm T} + i \rho^{\prime\prime}_{\rm T} &=& 1/ [ 
\lambda^\prime_{\rm T} + i \lambda^{\prime\prime}_{\rm T} ] \ .  \label{D07}
\end{eqnarray}
Clearly, Eqs.\ (\ref{D01}) and (\ref{D05}) indicate that the heat current 
${\bf Q}$ and the temperature gradient $\bbox{\nabla}T$ are not parallel
to each other. This fact is the consequence of the symmetry breaking by the
rotation frequency ${\bf \Omega}=\Omega {\bf e}_z$. In terms of the components
the heat transport equation (\ref{D01}) can be written as
\begin{eqnarray}
Q_x + iQ_y &=& - (\lambda^\prime_{\rm T} + i \lambda^{\prime\prime}_{\rm T} )
(\partial_x T + i\partial_y T) \ ,  \label{D08}  \\
Q_z &=& - \lambda_{\rm T,1} \partial_z T \ .  \label{D09}
\end{eqnarray}
Analogously, Eq.\ (\ref{D05})  can be written as
\begin{eqnarray}
\partial_x T + i\partial_y T &=& - (\rho^\prime_{\rm T} + 
i \rho^{\prime\prime}_{\rm T} ) (Q_x + iQ_y) \ ,  \label{D10}  \\
\partial_z T &=& - \rho_{\rm T,1} Q_z \ .  \label{D11}
\end{eqnarray}
Thus, the heat transport decouples into two independent contributions,
parallel and perpendicular to the rotation axis. The parallel heat transport
(in the $z$ direction) is very similar to the heat transport in nonrotating
$^4$He: the thermal conductivity $\lambda_{\rm T,1}$ defined in (\ref{D02})
has the same structure with the same amplitude $A_1$ as in the nonrotating
case (compare (\ref{D02}) and (\ref{B30}) with (5.2) and (4.33) in the second
paper of Ref.\ \onlinecite{H1}, respectively). The only difference is the 
function ${\cal F}_p(\zeta)$ which here is defined in a different way than in 
Ref.\ \onlinecite{H1}. On the other hand, the perpendicular heat transport
(in the $xy$ plane) shows a completely different nature. In this cases the
thermal conductivity $\lambda^\prime_{\rm T}+ i\lambda^{\prime\prime}_{\rm T}$ 
and the thermal resistivity $\rho^\prime_{\rm T}+i\rho^{\prime\prime}_{\rm T}$ 
are complex. While the real parts $\lambda^\prime_{\rm T}$ and 
$\rho^\prime_{\rm T}$ describe the dissipation of the heat current, 
the imaginary parts $\lambda^{\prime\prime}_{\rm T}$ and 
$\rho^{\prime\prime}_{\rm T}$ imply a coupling between the $x$ and $y$ 
direction so that the heat current and the related temperature gradient do not
have the same direction in the $xy$ plane. The real part $\rho^\prime_{\rm T}$
may be interpreted as the {\it longitudinal} thermal resistivity, because for
a given heat current in the $xy$ plane it implies an antiparallel temperature 
gradient. On the other hand, the imaginary part $\rho^{\prime\prime}_{\rm T}$
may be interpreted as the {\it transversal} thermal resistivity, because it
implies a temperature gradient perpendicular to the heat current. The 
amplitudes $A^\prime$ and $A^{\prime\prime}$ defined in (\ref{B55}) and
(\ref{B56}) depend on completely different integrals than $A_1$, which are
defined in (\ref{B38})-(\ref{B40}) and in (\ref{B26}), respectively, and 
depend strongly on the rotation frequency $\Omega$. 

The special structures of the conductivities and resistivities reflect the
fact that in rotating $^4$He there are straight vortex lines present which
move and imply dissipation for a perpendicular heat flow but do not move for
a parallel heat flow. The existence of a temperature gradient perpendicular
to the heat current is the analogy of the Hall effect in electronic systems,
where here the rotation frequency $\Omega$ plays the role of the magnetic
field. 

\begin{figure}[t]
\vspace*{6.9cm}
\includegraphics{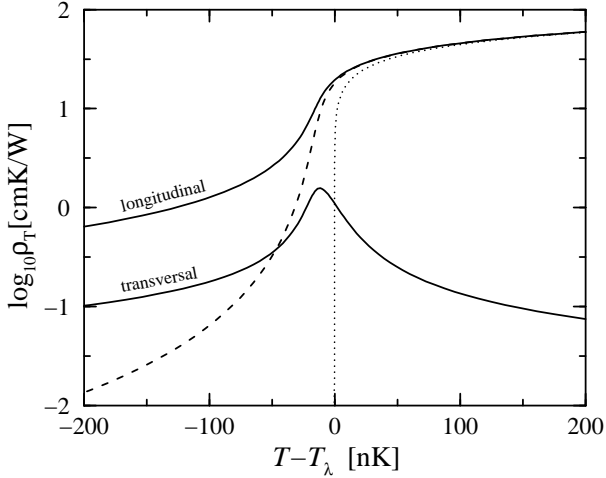}
\caption{The several thermal resistivities of rotating $^4$He as functions
of temperature for $\Omega=2\pi\ {\rm s^{-1}}$. The longitudinal resistivity 
$\rho^\prime_{\rm T}$ and the transversal resistivity 
$\rho^{\prime\prime}_{\rm T}$ for heat transport perpendicular to the 
rotation axis are shown as solid lines. The dashed line represents the 
resistivity $\rho_{\rm T,1}$ for heat transport parallel to the rotation axis. 
For comparison, the resistivity for $\Omega=0$ is shown as dotted line.}
\label{Fig04}
\end{figure}

We calculate the thermal conductivities numerically by (\ref{D02})-(\ref{D04})
for $d=3$ where the effective parameters and their gradients determined in 
Sec.\ \ref{S02} are inserted. From (\ref{D06}) and (\ref{D07}) we obtain 
the related resistivities. In Fig.\ \ref{Fig04} our results for the thermal 
resistivities are shown for $\Omega= 2\pi\ {\rm s^{-1}}$. The solid lines 
represent the longitudinal resistivity $\rho^\prime_{\rm T}$ and the 
transversal resistivity $\rho^{\prime\prime}_{\rm T}$ for the perpendicular 
heat flow, while the dashed line represents the resistivity $\rho_{\rm T,1}$ 
for the parallel heat flow. For comparison, the thermal resistivity for 
$\Omega=0$ is shown as dotted line. Clearly, for nonzero $\Omega$ the curves 
are smooth near $T_\lambda$. While $\rho_{\rm T,1}$ and $\rho^\prime_{\rm T}$ 
increase continuously with increasing temperature, the transversal 
conductivity $\rho^{\prime\prime}_{\rm T}$ exhibits a maximum near
$T_\lambda$. The inflection points of $\log_{10}\rho_{\rm T,1}$ and $\log_{10} 
\rho^\prime_{\rm T}$ and the maximum of $\log_{10}\rho^{\prime\prime}_{\rm T}$ 
in Fig.\ \ref{Fig04} are located at a temperature below $T_\lambda$ close to 
the temperature of the maximum of the specific heat in Fig.\ \ref{Fig02}.
Thus, the shift $\Delta T_\lambda(\Omega)$ of the superfluid transition
temperature for nonzero $\Omega$, defined in (\ref{C05}), is observed also in 
the thermal conductivities where here the constant $M_\lambda$ is slightly 
different.

For temperatures $T$ well above and well below $T_\lambda$ asymptotic 
formulas can be derived for the thermal conductivities and resistivities.
To do this we must evaluate the integrals ${\cal F}_p(\zeta)$, ${\cal M}_p
(\zeta)$, ${\cal N}^\prime_p(\zeta)$, and ${\cal N}^{\prime\prime}_p(\zeta)$ 
for $\zeta\gg +1$ and for $\zeta\rightarrow -1$, respectively. While for
${\cal F}_p(\zeta)$ the asymptotic formula is given by (\ref{B61}), analogous
asymptotic formulas are found for the other integrals. 

First, we consider $T\gg T_\lambda$ where $\zeta=r_1\ell^2\gg +1$. In this 
case the flow-parameter condition (\ref{B60}) reduces to $r_1\xi^2=1$.
Then, for the amplitudes (\ref{B53})-(\ref{B56}) we obtain $A\approx 0$,
$A_1 \approx A^\prime \approx -1/2$, and $A^{\prime\prime} \approx 0$.
Consequently, from (\ref{D06}) and (\ref{D07}) together with 
(\ref{D02})-(\ref{D04}) we obtain the thermal resistivities
\begin{equation}
\rho_{\rm T,1} \approx \rho^\prime_{\rm T} \approx {\tau \, \xi^{d-2}\over 
g_0 k_{\rm B} A_d} \, {2 \gamma[\tau] F[\tau] \over \{ 1 - f[\tau]/4 \} }  
\label{D12}
\end{equation}
and $\rho^{\prime\prime}_{\rm T}\approx 0$ where $\xi=\xi_0 \tau^{-\nu}$ and 
$\tau=(T-T_\lambda)/T_\lambda$. Because of (\ref{B57}) and (\ref{B63}) the 
RG flow parameter $\tau$ is identified by the reduced temperature. 
Eq.\ (\ref{D12}) is the well known formula for the thermal resistivity for 
$\Omega=0$ and $Q\rightarrow 0$ with the amplitude functions evaluated up to 
one-loop order (see Ref.\ \onlinecite{D1}). In Fig.\ \ref{Fig04} this 
resistivity is shown by the dotted line. Thus, for increasing temperatures 
$T$ well above $T_\lambda$ the thermal resistivities $\rho_{\rm T,1}$ and 
$\rho^\prime_{\rm T}$ asymptotically approach the resistivity for $\Omega=0$
where the imaginary part $\rho^{\prime\prime}_{\rm T}$ decreases and is 
negligibly small. This fact is clearly seen in Fig.\ \ref{Fig04}. 

Secondly, we consider $T\ll T_\lambda$ where $\zeta=r_1\ell^2 \rightarrow -1$. 
In this case in the flow-parameter condition (\ref{B60}) the first term can
be neglected so that
\begin{equation}
r_1\ell^2 +1 = \Bigl\{ {32 u[\tau]\over\epsilon} (1-\epsilon/2) 
(\xi/\ell)^{2-\epsilon} \Bigr\}^{2/\epsilon} 
\label{D13}
\end{equation}
which is very small for $\xi\ll\ell$. Because of (\ref{B57}) and (\ref{B64})
the RG flow parameter is related to the reduced temperature by $\tau=
2(T_\lambda-T)/T_\lambda$. Inserting the asymptotic formulas of the integrals
(\ref{B26}) and (\ref{B38})-(\ref{B40}) for $\zeta\rightarrow -1$ into
(\ref{B53})-(\ref{B56}) we obtain the amplitudes $A$, $A_1$, $A^\prime$,
and $A^{\prime\prime}$. Then, from (\ref{D02})-(\ref{D04}) we obtain the 
thermal conductivities, and from (\ref{D06}) and (\ref{D07}) we obtain the
related resistivities. Eventually, in leading order we obtain the asymptotic
formulas
\begin{eqnarray}
\rho_{\rm T,1} &\approx& (4/\epsilon)(r_1\ell^2+1) \, \rho^\prime_{\rm T}  \ , 
\label{D14}  \\
\rho^\prime_{\rm T} &\approx& {\tau \, \xi^{d-2} \over g_0 k_{\rm B} A_d} 
\, {4\gamma[\tau] w^\prime[\tau] \over F[\tau] } \, 8u[\tau]
\, (\xi/\ell)^2  \label{D15}  \\
\rho^{\prime\prime}_{\rm T} &\approx& {\tau \, \xi^{d-2} \over g_0 k_{\rm B} 
A_d} \, {4\gamma[\tau] w^{\prime\prime}[\tau] \over F[\tau] } \, 8u[\tau] 
\, (\xi/\ell)^2  \ ,  \label{D16}
\end{eqnarray}
where $\xi=\xi_0\tau^{-\nu}$ and $\ell=(\hbar/2m_4\Omega)^{1/2}$. 

For the heat transport perpendicular to the rotation axis for $T\ll T_\lambda$
the transversal and the longitudinal resistivities are related to each other by
\begin{equation}
\rho^{\prime\prime}_{\rm T} / \rho^\prime_{\rm T} \approx 
w^{\prime\prime}[\tau] / w^\prime[\tau]  \ . 
\label{D17}
\end{equation}
 Neglecting the nonasymptotic effects due to the renormalized coupling 
 parameters $\gamma[\tau]$, $w^\prime[\tau]$, $w^{\prime\prime}[\tau]$, and
 $F[\tau]$, we obtain asymptotically in the leading order 
\begin{equation}
\rho^\prime_{\rm T} \sim \rho^{\prime\prime}_{\rm T} \sim \tau\,\xi^d 
\ell^{-2} \sim \Omega\, (T_\lambda-T)^{-1+\alpha} 
\label{D18}
\end{equation}
where $\alpha=2-d\nu = -0.013$. Thus, in slowly rotating superfluid $^4$He
the thermal resistivities for the perpendicular heat flow depend linearly on
the rotation frequency. Since the resistivities are caused by vortex lines, 
this result is plausible: the resistivities must be proportional to the
total length of the vortex lines per unit volume $L$ which is related to the
rotation frequency by $L=1/2\pi\ell^2 = (m_4/\pi\hbar)\Omega$. For $\Omega
\rightarrow 0$ the thermal resistivities $\rho^\prime_{\rm T}$ and 
$\rho^{\prime\prime}_{\rm T}$ vanish as expected. 

On the other hand, the heat flow parallel to the rotation axis is not
directly influenced by the vortex lines, because the heat current flows 
parallel to the vortex lines. The dissipation of a parallel heat current and 
the related resistivity $\rho_{\rm T,1}$ must be caused by fluctuation 
effects. For this reason $\rho_{\rm T,1}$ is much smaller than the resistivity
$\rho^\prime_{\rm T}$ of the perpendicular heat flow. This fact is clearly 
seen in Fig.\ \ref{Fig04} and in Eq.\ (\ref{D14}) because the prefactor 
defined by (\ref{D13}) is very small. Neglecting the nonasymptotic effects
due to the renormalized coupling parameters, we obtain asymptotically in 
leading order 
\begin{equation}
\rho_{\rm T,1} \sim \tau\,\xi^{d-2} (\xi/\ell)^{4/\epsilon} = 
\tau^{1-(d-2+4/\epsilon)\nu} \ell^{-4/\epsilon}  \ . 
\label{D19}
\end{equation}
For $d=3$ dimensions where $\epsilon=4-d=1$ we find $\rho_{\rm T,1}\sim 
\Omega^2 (T_\lambda -T)^{1-5\nu}$, so that the resistivity for the parallel 
heat flow is quadratical in the rotation frequency. Again, $\rho_{\rm T,1}$ 
vanishes for $\Omega\rightarrow 0$ as expected. The temperature dependence 
of $\rho_{\rm T,1}$ is very similar to the temperature dependence of the
thermal resistivity in nonrotating $^4$He for nonzero heat currents $Q$
(compare the dashed line in Fig.\ \ref{Fig04} with the solid lines in
Figs.\ 3 and 4 in the second paper of Ref.\ \onlinecite{H1}). The related 
exponents for the approximate power laws for $T\ll T_\lambda$ are nearly 
the same.

\section{Mutual friction and \break the Vinen coefficients} \label{S06}
Mutual friction in rotating superfluid $^4$He was investigated first by Hall
and Vinen \cite{HV}. While originally the attenuation of second sound in
resonant cavities was considered \cite{HV}, the mutual friction implies
dissipation and a related resistance for the heat transport in the superfluid
state. Here we show that the phenomenological theory of Hall and Vinen 
implies a heat-transport equation which has the same form as (\ref{D05}) 
obtained within our approach. In this way we show that mutual friction 
according to Hall and Vinen is derived from model {\it F\,} by our approach.

Hall and Vinen proposed a mutual-friction force between the superfluid and 
the normal-fluid component which is implied by the motion of the vortex lines 
in the presence of a superfluid-normal-fluid counterflow. This mutual-friction 
force must be added to and subtracted from the two-fluid hydrodynamic 
equations for the superfluid and the normal-fluid component, respectively. 
For a stationary homogeneous counterflow with the relative velocity 
${\bf v}_{\rm s}-{\bf v}_{\rm n}$ a relation for the temperature gradient 
$\bbox{\nabla}T$ can be derived from the two-fluid hydrodynamic equations 
which reads 
\begin{eqnarray}
(\rho/\rho_{\rm n})\, s \bbox{\nabla}T&=& -B\Omega\, ({\bf e}_z\times 
({\bf e}_z\times ({\bf v}_{\rm s}-{\bf v}_{\rm n})) \nonumber\\
&&+ (2-B^\prime) \Omega\, ({\bf e}_z \times ({\bf v}_{\rm s}-{\bf v}_{\rm n})) 
\ .  \label{E01}
\end{eqnarray}
Here, $B$ and $B^\prime$ are the Vinen coefficients which represent the
phenomenological parameters of the mutual-friction force. The term with the
coefficient $2$ arises from the Coriolis force in the rotating frame. In the
critical regime close to $T_\lambda$ it is $\rho/\rho_{\rm n}\approx 1$ and
the entropy per mass is $s\approx s_\lambda= (\hbar/m_4)(g_0/T_\lambda)$ 
where $g_0=2.164 \times 10^{11}\ {\rm s^{-1}}$ at saturated vapor pressure
\cite{TA}.

For a zero net mass current the superfluid-normal-fluid counterflow with 
relative velocity ${\bf v}_{\rm s}-{\bf v}_{\rm n}$ is directly related to
the heat current ${\bf Q}$. Considering the counterflow as a metastable state,
in Ref.\ \onlinecite{HD2} the heat current ${\bf Q}$ was calculated for 
nonrotating superfluid $^4$He as a function of the counterflow wave vector 
${\bf k}=(m_4/\hbar)({\bf v}_{\rm s}-{\bf v}_{\rm n})$. The calculation was
performed in the critical regime near $T_\lambda$ using model {\it F\,} and
the RG theory. Since the rotation frequency $\Omega$ is very small, we may 
use this result here. For small counterflows and small heat currents the 
relation between ${\bf Q}$ and ${\bf v}_{\rm s}-{\bf v}_{\rm n}$ is linear. 
While in Ref.\ \onlinecite{HD2} this relation was calculated in one-loop RG 
theory, for our purpose the relation in zero-loop RG theory is sufficient 
which reads
\begin{equation}
{\bf Q}= - {g_0 k_{\rm B} T_\lambda \over \xi^{d-2}} \, {A_d\over 8u[\tau]}
\, {m_4\over \hbar}\, ({\bf v}_{\rm s}-{\bf v}_{\rm n})  \ . 
\label{E02}
\end{equation}
Now, we solve (\ref{E02}) for ${\bf v}_{\rm s}-{\bf v}_{\rm n}$ and insert 
the resulting expression into (\ref{E01}). Then, we obtain the temperature 
gradient 
\begin{eqnarray}
\bbox{\nabla}T &=& {\xi^{d-2}\over g_0^2 k_{\rm B} A_d} \, 8u[\tau]\, B\Omega 
\, ({\bf e}_z\times ({\bf e}_z\times {\bf Q})) \nonumber\\
&&- {\xi^{d-2}\over g_0^2 k_{\rm B} A_d} \, 8u[\tau]\, (2-B^\prime) \Omega 
\, ({\bf e}_z \times {\bf Q})  \ . 
\label{E03}
\end{eqnarray}
This formula has the same structure as (\ref{D05}). By comparison we find the
thermal resistivities
\begin{eqnarray}
\rho^\prime_{\rm T} &=& {\xi^{d-2}\over g_0^2 k_{\rm B} A_d} \, 8u[\tau]\, 
B\Omega  \ ,  \label{E04}  \\
\rho^{\prime\prime}_{\rm T} &=& {\xi^{d-2}\over g_0^2 k_{\rm B} A_d} \, 
8u[\tau]\, (2-B^\prime) \Omega  \ ,  \label{E05}
\end{eqnarray}
and $\rho_{\rm T,1}=0$. For the heat flow perpendicular to the rotation axis
the resistivities (\ref{E04}) and (\ref{E05}) are a direct consequence of the
mutual friction due to the motion of the vortices. Because of $\Omega=
(\pi\hbar/m_4)L$ these resistivities are proportional to the vortex density 
$L$ which precisely is the total length of the vortex lines per volume. 
Since $\rho^\prime_{\rm T}$ implies a temperature gradient antiparallel to
the heat current, the Vinen coefficient $B$ describes the dissipative effects
of the vortex lines. On the other hand, $\rho^{\prime\prime}_{\rm T}$ implies
a temperature gradient perpendicular to the heat current, so that the Vinen
coefficient $B^\prime$ and the Coriolis-force coefficient $2$ are related to
reversible effects. For a heat flow parallel to the rotation axis the vortex
lines do not move and hence do not cause dissipation, so that the resistivity
$\rho_{\rm T,1}$ is zero. 

Within our approach which is based on the Hartree approximation combined with
the RG theory we have derived the formula (\ref{D05}) for the temperature
gradient. The terms involving the heat current perpendicular to the rotation
axis have the same structure as the terms on the right-hand side of 
(\ref{E03}). Thus, we conclude that our approach includes the mutual friction 
between the superfluid and the normal-fluid component caused by the motion
of vortices. However, since the superfluid state is homogeneous in space
and vortex lines are not treated explicitly, our approach includes the effect
of the vortices {\it indirectly}. Furthermore, our approach includes also
fluctuation effects, which is reflected by the first term on the right-hand
side of (\ref{D05}) and the nonzero resistivity $\rho_{\rm T,1}$ for the 
parallel heat flow. This term is not present in (\ref{E03}).

We determine the Vinen coefficients by solving (\ref{E04}) and (\ref{E05}) 
for $B$ and $2-B^\prime$ and inserting the thermal resistivities 
$\rho^\prime_{\rm T}$ and $\rho^{\prime\prime}_{\rm T}$ obtained within our
approach. For small rotation frequencies $\Omega$ the thermal resistivities 
are given by the asymptotic formulas (\ref{D15}) and (\ref{D16}). Thus, we 
obtain the Vinen coefficients 
\begin{eqnarray}
B &=& (2m_4/\hbar) \, g_0 \tau\, \xi^2 \, 4\gamma[\tau] w^\prime[\tau]
/F[\tau]  \ ,  \label{E06}  \\
2-B^\prime &=& (2m_4/\hbar) \, g_0 \tau\, \xi^2 \, 4\gamma[\tau] 
w^{\prime\prime}[\tau]/F[\tau]  \ ,  \label{E07}
\end{eqnarray}
which are defined in the limit $\Omega\rightarrow 0$. These formulas are
well suited for numerical evaluations because all parameters and constants
are known. The RG flow parameter is related to the reduced temperature by
$\tau=2(T_\lambda-T)/T_\lambda$. 

Simpler expressions for the Vinen coefficients are obtained if we replace 
$g_0$ by the renormalized counterpart $g(l)=g[\tau]$. From the 
renormalization equation (\ref{B32}) we obtain
\begin{equation}
g_0 = g(l)\, (\chi_0 Z_m(l))^{1/2} ((\mu l)^\epsilon/A_d)^{1/2} \ .  
\label{E08}
\end{equation}
We solve (\ref{B49}) for $(\chi_0 Z_m(l))^{1/2}$ and insert the
resulting expression into (\ref{E08}). Then we obtain
\begin{equation}
g_0 = g[\tau] / (\tau\, \xi^2 2\gamma[\tau])
\label{E09}
\end{equation}
where $\mu l=\xi^{-1}$ has been identified. Now, we insert this result for
$g_0$ into (\ref{E06}) and (\ref{E07}). Furthermore, we insert the 
dimensionless renormalized couplings $w^\prime[\tau]=\Gamma^\prime[\tau]/
\lambda[\tau]$, $w^{\prime\prime}[\tau]=\Gamma^{\prime\prime}[\tau]/\lambda
[\tau]$, and $F[\tau]=g[\tau]/\lambda[\tau]$. Since most renormalized 
couplings cancel, we obtain eventually 
\begin{eqnarray}
B &=& (4m_4/\hbar) \, \Gamma^\prime[\tau]  \ ,  \label{E10}  \\
2-B^\prime &=& (4m_4/\hbar) \, \Gamma^{\prime\prime}[\tau]  \ .  \label{E11}
\end{eqnarray}
This result is remarkably simple. Eqs.\ (\ref{E10}) and (\ref{E11}) represent
the final formulas for the Vinen coefficients obtained within our approach.

The Vinen coefficients were calculated previously for model {\it F\,} within
the renormalized mean-field theory \cite{H2}. In this previous approach the 
the vortex lines are considered explicitly. The model-{\it F\,} equations are 
written first in the renormalized form and then solved as mean-field 
equations. Solutions were obtained which represent a single straight vortex
line moving under the influence of a superfluid-normal-fluid counterflow. 
From the relation between the velocity of the vortex line ${\bf v}_{\rm L}$ 
and the relative velocity of the counterflow ${\bf v}_{\rm s}-{\bf v}_{\rm n}$ 
the Vinen coefficients $B$ and $2-B^\prime$ were extracted. Eventually, the 
effects of the critical fluctuations are included by application of the RG 
theory. The theoretical results were compared with the experimental data of 
Refs.\ \onlinecite{SL,L1,MS} obtained for temperatures $T_\lambda-T \gtrsim 
3\times 10^{-4}\ {\rm K}$. For $B$ the agreement between the experiments and 
the previous theory \cite{H2} is very good, while for $2-B^\prime$ there are 
some discrepancies. Consequently, the results of the previous theory may be 
viewed as correct and reliable. 

Close to criticality, the leading temperature dependence of the Vinen 
coefficients obtained from the previous theory is governed by
\begin{eqnarray}
B + i(2-B^\prime) &\approx& (2m_4/\hbar) \Gamma^\prime[\tau] /[\, c_1 + c_2 
(\gamma[\tau] F[\tau])^2  \nonumber\\
&&\hskip2cm - i\, c_3 \gamma[\tau] F[\tau]\,] 
\label{E12}
\end{eqnarray}
where $c_1=0.2511$, $c_2=0.391$, and $c_3=0.795$ (see Eq.\ (37) and Table 1 
in Ref.\ \onlinecite{H2}, the winding number of the vortices is assumed to be
$n=1$). Since the product of the renormalized couplings $\gamma[\tau] F[\tau]$ 
is small, the term with the coefficient $c_2$ can be neglected so that 
(\ref{E12}) simplifies into
\begin{eqnarray}
B &\approx& (4m_4/\hbar) \Gamma^\prime[\tau] \times 1.99  \ , \label{E13} \\
2-B^\prime &\approx& (4m_4/\hbar) \Gamma^\prime[\tau] \gamma[\tau] F[\tau]
\times 6.30  \ .  \label{E14}
\end{eqnarray}
These formulas must be compared with (\ref{E10}) and (\ref{E11}). For $B$
the leading critical temperature dependence is governed by the renormalized 
coupling $\Gamma^\prime[\tau]$ in both cases, in the present theory and in the 
previous theory \cite{H2}. Quantitatively, the previous theory predicts a 
$1.99$ times larger result for $B$ than our present theory. On the other hand, 
for $2-B^\prime$ the leading critical temperature dependence is governed by 
different renormalized couplings for the two approaches, which is clearly 
seen in (\ref{E11}) and (\ref{E14}). The critical divergence of 
$\Gamma^{\prime\prime}[\tau]$ is much weaker than the critical divergences 
of $\Gamma^\prime[\tau]$ and of $\Gamma^\prime[\tau] \gamma[\tau] F[\tau]$. 

If we generalize model {\it F\,} by replacing the complex order parameter 
$\psi$ by a complex vector $\Psi=(\psi_1,\ldots,\psi_n)$ of $n$ components, it 
turns out that the Hartree approximation is exact in the limit $n\rightarrow
\infty$. The RG theory can be applied but is not necessary in this case. 
Furthermore, in the limit $\Omega\rightarrow 0$ for $T<T_\lambda$ the 
approximation reduces into the mean-field theory. Consequently, 
Eqs.\ (\ref{E10}) and (\ref{E11}) may be interpreted as the Vinen coefficients
for $n=\infty$ in renormalized mean-field theory. On the other hand, 
previously \cite{H2} we have calculated the Vinen coefficients for $n=1$ in 
renormalized mean-field theory, which are given by (\ref{E13}) and (\ref{E14}) 
in leading order close to criticality. Thus, the Vinen coefficients allow a 
direct comparison of the physics for $n=1$ and $n=\infty$ where the same 
approximation scheme is applied. While for $B$ the agreement is quite good,
for $2-B^\prime$ we find serious discrepances.

\section{Comparison with the Gorter- \break Mellink mutual friction in \break
nonrotating $^4$He at nonzero \break heat currents} \label{S07}
Our approach was originally developed in Ref.\ \onlinecite{H1} for nonrotating 
$^4$He in the presence of a nonzero heat current $Q$. For $T<T_\lambda$ we 
found mutual friction between the superfluid and the normal-fluid component 
according to Gorter and Mellink \cite{GM} and calculated the Gorter-Mellink 
coefficient $A$. In the superfluid region we found a nonzero thermal 
resistivity which for temperatures $T$ well below $T_\lambda$ is given by the 
asymptotic formula \cite{H1}
\begin{equation}
\rho_{\rm T} \approx {\tau\, \xi^{d-2} \over g_0 k_{\rm B}} \,
{4\gamma[\tau] w^\prime[\tau] \over F[\tau]} \, 
\sqrt{12} \, \Bigl( {8u[\tau] \over (-\zeta)^{1/2} A_d} \Bigr)^3
\Bigl( {Q\, \xi^{d-1} \over g_0 k_{\rm B} T_\lambda} \Bigr)^2 
\label{F01}
\end{equation}
where the RG flow parameter $\tau$ is related to the reduced temperature by
$\tau=2(T_\lambda-T)/T_\lambda$. Furthermore, we found a second correlation
length $\xi_1$ which is a dephasing length for the order-parameter field. For
$T\ll T_\lambda$ we obtain the asymptotic formula \cite{H1} 
\begin{equation}
\xi_1 \approx \bigl[{\textstyle{4\over 3}} (-\zeta)^3 \bigr]^{1/4} 
{A_d\over 8u[\tau]} \ {g_0 k_{\rm B} T_\lambda \over Q\,\xi^{d-2} } \ . 
\label{F02}
\end{equation}
Here $\zeta$ is a dimensionless variable which depends logarithmically on the
reduced temperature and varies between $-5$ and $-15$ in the superfluid 
region. $\zeta$ can be eliminated in favor of the dephasing length $\xi_1$. 
To do this, we solve (\ref{F02}) for $\zeta$ and insert the resulting 
expression into (\ref{F01}). Then, we obtain the thermal resistivity
\begin{equation}
\rho_{\rm T} \approx {\tau \, \xi^{d-2} \over g_0 k_{\rm B} A_d} 
\, {4\gamma[\tau] w^\prime[\tau] \over F[\tau] } \, 8u[\tau] 
\, (2\xi/\xi_1)^2  \ . 
\label{F03}
\end{equation}

Vinen argued \cite{V1} that in superfluid $^4$He a nonzero heat current $Q$
implies a turbulent superfluid flow and a tangle of vortex lines. The mutual
friction and the dissipation of the heat current are caused by the motion of
the vortex lines. The thermal resistivity (\ref{F03}) has nearly the same
structure as $\rho^\prime_{\rm T}$ in (\ref{D15}) where here the dephasing
length $\xi_1$ plays the role of the characteristic length $\ell$. Thus, the
resistivity (\ref{F03}) must be due to dissipation by the motion of vortex
lines, which agrees with Vinen's picture \cite{V1}. 

The thermal resistivities (\ref{D14})-(\ref{D16}) represent the elements of 
the resistivity tensor for a heat flow in the presence of uniformly 
distributed straight vortex lines. For a tangle of vortex lines the thermal 
resistivity is obtained by taking the average over all directions of the
vortex lines. Thus, we obtain
\begin{equation}
\rho_{\rm T} \approx {\textstyle{1\over 3}} ( \rho^\prime_{\rm T} + 
\rho^\prime_{\rm T} + \rho_{\rm T,1} ) \approx {\textstyle{2\over 3}}
\rho^\prime_{\rm T}
\label{F04}
\end{equation}
which is just the trace of the resistivity tensor divided by $d=3$. Because
of $\rho_{\rm T,1}\ll \rho^\prime_{\rm T}$, the resistivity $\rho_{\rm T,1}$ 
of the heat flow parallel to the vortex lines can be neglected. Now, 
inserting (\ref{D15}) into (\ref{F04}) we obtain
\begin{equation}
\rho_{\rm T} \approx {2\over 3}\, {\tau \, \xi^{d-2} \over g_0 k_{\rm B} A_d} 
\, {4\gamma[\tau] w^\prime[\tau] \over F[\tau] } \, 8u[\tau] 
\, (\xi/\ell)^2  \ . 
\label{F05}
\end{equation}
The prefactor $2/3$ is due to the assumption that the direction of the vortex
lines is isotropically distributed. However, since the isotropy is broken
by a homogeneous heat current, the assumption may not be perfectly true.
Thus, the correct prefactor will be slightly different from $2/3$.

Now, we compare (\ref{F05}) with (\ref{F03}). We eliminate $\rho_{\rm T}$ and
solve the resulting equation for $\ell^{-2}$. Then, we obtain the total 
length of the vortex lines per volume
\begin{equation}
L = 1/(2\pi\ell^2) = (3/\pi)\, \xi_1^{-2} 
\label{F06}
\end{equation}
which is the density of the vortex lines induced by the heat current $Q$. This
result indicates that the dephasing length $\xi_1$ can be interpreted as the
mean distance between the vortex lines. Inserting (\ref{F02}) for the 
dephasing length we find in leading order
\begin{equation}
L \sim Q^2 \sim ({\bf v}_{\rm s} - {\bf v}_{\rm n})^2 
\label{F07}
\end{equation}
so that the vortex density increases with the square of the counterflow 
velocity. Eq.\ (\ref{F07}) agrees with the vortex density proposed by Vinen
\cite{V1} and with the ansatz of Gorter and Mellink \cite{GM} for the 
mutual-friction force. Furthermore, in leading order we find that the 
temperature dependence of the vortex density is given by $L\sim 
(T_\lambda-T)^{2(d-2)\nu}$.

The average thermal resistivity (\ref{F04}) of the vortex tangle can be 
expressed in terms of the Vinen coefficient $B$ and the vortex density $L$. 
For this purpose we insert (\ref{E04}) into (\ref{F04}), replace $\Omega=
(\pi\hbar/m_4)L$, and obtain
\begin{equation}
\rho_{\rm T} = {2\over 3}\, {\xi^{d-2}\over g_0^2 k_{\rm B} A_d} 
\, 8u[\tau]\, {\pi\hbar\over m_4}\, B\, L  \ ,  
\label{F08}
\end{equation}
i.e.\ $\rho_{\rm T}\sim BL$. Thus, there are two sources which determine the
magnitude of the thermal resistivity: the dissipation effects of a single
vortex line described by $B$ and the density of the vortices $L$. 

In Ref.\ \onlinecite{H1} we found that for a homogeneous heat current $Q$ in
superfluid $^4$He our approach considerably overestimates the dissipation by 
the vortex lines.
Our theory predicts a Gorter-Mellink coefficient $A$ which is about $20$ 
times larger than observed in the experiments \cite{GM,V1}. This disagreement 
was confirmed recently by a measurement \cite{BA} of the thermal resistivity 
$\rho_{\rm T}$ in superfluid $^4$He close to $T_\lambda$. We may now ask the 
question whether this large discrepancy arises due to the dissipation of the 
single vortex lines described by $B$ or by the vortex density $L$. Rotating 
superfluid $^4$He is the well suited system for a test of our approach to 
answer this question. Since the vortex density $L=(m_4/\pi\hbar)\Omega$ is 
fixed and precisely known for this system, any observed discrepancy of the 
dissipation in rotating superfluid $^4$He must be due to the Vinen coefficient 
$B$. In Sec.\ \ref{S06} we have compared our present result for $B$ with the 
result of the previous theory \cite{H2}. While our present approach yields the 
correct critical temperature dependence, the values of $B$ are smaller by a 
factor of $2$ than expected. Thus, for a single vortex line our approach 
predicts a two times smaller dissipation than expected. On the other hand, 
for the heat transport at nonzero currents $Q$ our approach predicts a $20$ 
times larger dissipation. This large discrepancy cannot be caused by the 
Vinen coefficient $B$. Rather, Eq.\ (\ref{F08}) implies that in 
Ref.\ \onlinecite{H1} our approach predicts a $40$ times larger vortex 
density $L$. Thus, we conclude that for superfluid $^4$He in the presence of 
a heat current $Q$ our theory considerably overestimates the density of the 
vortex lines $L$, while for the dissipative effect of a single vortex line 
described by $B$ the correct order of magnitude is obtained. 

The discrepancy may possibly be explained in the following way. The Hartree
approximation implies that our approach effectively considers the generalized
model {\it F\,} with $n=\infty$ complex order parameters. On the other hand,
real liquid $^4$He is described by model {\it F\,} with $n=1$ complex order
parameter. Thus, the discrepancy may be related to the difference of the
physics between $n=\infty$ and $n=1$. For $n=1$ a homogeneous heat current
$Q$ and the related superfluid-normal-fluid counterflow represent a metastable
state, which relaxes only by creation of vortices. For this purpose, energy
barriers must be overcome, so that the vortex-creation rate, the vortex 
density $L$, and hence the dissipation are strongly suppressed. On the other
hand for $n=\infty$ the heat current $Q$ is unstable and vortices are created
without an energy barrier. Consequently, in this case the dissipation, the
vortex-creation rate, and the vortex density $L$ are not suppressed. Thus, the
dissipation of the heat current and the vortex density may be considerably
larger for $n=\infty$ (our theoretical approach) than for $n=1$ (real 
superfluid $^4$He). The difference may be a large factor because the energy
barriers occur in the argument of an exponential function. Thus, the 
quantitative discrepancy by a factor of about $20$ or $40$ can be explained 
in this way.

\section{Conclusions} \label{S08}
Our recently developed approach \cite{H1}, the Hartree approximation 
combined with the renormalization-group theory for model {\it F}, can 
be applied also to rotating $^4$He close to the superfluid transition.
For nonzero rotation frequencies $\Omega$ our theory predicts that all 
physical quantities are smooth and round near $T_\lambda$. The superfluid 
transition is a smooth crossover located at a temperature 
$T_\lambda(\Omega)$ which is not sharply defined and shifted to lower 
temperatures by $\Delta T_\lambda(\Omega)=T_\lambda(\Omega)-T_\lambda= 
-M_\lambda T_\lambda(2m_4\xi_0^2\Omega/\hbar)^{1/2\nu}$ where $M_\lambda
\approx 1.2$. For the rotation frequencies $\Omega\approx 2\pi\ {\rm s^{-1}}$ 
of realistic experiments \cite{HV,SL,L1,MS} we find $\Delta T_\lambda(\Omega) 
\approx -25\ {\rm nK}$ which is very small. Thus, to observe the influence
of the rotation on the superfluid transition, experiments must be performed
with a temperature resolution of a few nano Kelvins. Since on earth the
gravity does not allow this temperature resolution these experiments must
be performed in a microgravity environment in space.

In superfluid $^4$He for $T<T_\lambda$ the thermal resistivity and the 
dissipation of the heat current strongly depend on the direction of the heat
flow related to the rotation axis. For a perpendicular heat flow dissipation
is caused by the motion of the vortex lines in the superfluid-normal-fluid
counterflow. On the other hand, for a parallel heat flow the thermal 
resistivity is considerably smaller and caused by fluctuation effects. 

From the thermal resistivities of the perpendicular heat flow we extract the
Vinen coefficients $B$ and $B^\prime$ in the limit $\Omega\rightarrow 0$. 
In this way we find that model {\it F\,} includes the mutual friction between
the superfluid and the normal-fluid component, which originally was proposed
by Hall and Vinen \cite{HV} on phenomenological grounds. While our theory 
does not treat the vortices microscopically, we nevertheless obtain 
dissipation effects due to vortices. Consequently, our approach includes the 
vortices indirectly.

\acknowledgments
I would like to thank Prof.\ Dr.\ V.\ Dohm for discussions and for
comments on the manuscript.

\appendix\section*{Calculation of \break the Green's function}
The equal time Green's function $G({\bf r},{\bf r}^\prime)$ is defined by
(\ref{B08})-(\ref{B10}). For the renormalization we need additionally the 
relations $\Gamma_0=Z^{-1}_\Gamma \Gamma$ and $\psi= Z^{1/2}_\psi 
\psi_{\rm ren}$ where, however, in Hartree approximation it is \cite{H1}
$Z_\Gamma=1$ and $Z_\psi=1$. Thus, from (\ref{B08}) we obtain the renormalized
Green's function
\begin{equation}
G({\bf r},{\bf r}^\prime) = 4\Gamma^\prime \int_0^\infty d\alpha 
\ e^{\alpha A} \, e^{\alpha B}  \,\delta({\bf r}-{\bf r}^\prime)  \ . 
\label{A01}
\end{equation}
By using the relations (\ref{B31}), (\ref{B32}), and $\Delta r_0=Z_r\Delta r$,
in the operators (\ref{B09}) and (\ref{B10}) we replace the bare parameters by
the renormalized counterparts. All $Z$ factors cancel except $Z_\Gamma$ which, 
however, is unity here. Thus, from (\ref{B09}) and (\ref{B10}) we obtain
\begin{eqnarray}
A&=&- \bigl\{ \Gamma [r_1 - (\bbox{\nabla}-i{\bf k})^2] - i(g/2\gamma) 
\Delta r \bigr\} \ ,  \label{A02}  \\
B&=&- \bigl\{ \Gamma^* [r_1 - (\bbox{\nabla}-i{\bf k})^2] + i(g/2\gamma) 
\Delta r \bigr\} \ .  \label{A03}
\end{eqnarray}
In the following we must figure out, how the operators $e^{\alpha A}$ and
$e^{\alpha B}$ act on the delta function $\delta({\bf r}-{\bf r}^\prime)$
to evaluate the integrand in (\ref{A01}). 

We derive an explicit integral formula for the Green's function for rotating
$^4$He in the presence of an infinitesimal heat current ${\bf Q}$ and an 
infinitesimal temperature gradient $\bbox{\nabla}T$. For the effective 
parameters $r_1$ and $\Delta r$ we use the ansatz
\begin{eqnarray}
r_1 &=& a_1 + {\bf b}_1 {\bf r} \ ,  \label{A04} \\
\Delta r &=& a + {\bf b}\, {\bf r} \ ,  \label{A05}
\end{eqnarray}
where $a_1$, $a$ are constants and ${\bf b}_1$, ${\bf b}$ are infinitesimal 
vectors which are related to the temperature gradient $\bbox{\nabla}T$. We 
evaluate the Green's function (\ref{A01}) up to first order in ${\bf b}_1$ 
and ${\bf b}$. For this purpose we decompose 
\begin{eqnarray}
A &=& A_0 + \Delta A \ ,  \label{A06} \\
B &=& B_0 + \Delta B \ ,  \label{A07}
\end{eqnarray}
where 
\begin{eqnarray}
A_0&=&- \bigl\{ \Gamma [a_1 - (\bbox{\nabla}-i{\bf k})^2] - i(g/2\gamma) 
a \bigr\} \ ,  \label{A08}  \\
B_0&=&- \bigl\{ \Gamma^* [a_1 - (\bbox{\nabla}-i{\bf k})^2] + i(g/2\gamma) 
a \bigr\} \ ,  \label{A09}
\end{eqnarray}
and
\begin{eqnarray}
\Delta A&=&-[ \Gamma {\bf b}_1 - i(g/2\gamma) {\bf b}]\, {\bf r} \ ,
\label{A10}  \\
\Delta B&=&-[ \Gamma^* {\bf b}_1 + i(g/2\gamma) {\bf b}]\, {\bf r} \ .
\label{A11} 
\end{eqnarray}
We expand the operator $e^{\alpha A}=e^{\alpha A_0 + \alpha\Delta A}$ up to
first order in $\Delta A$ according to 
\begin{eqnarray}
e^{\alpha A} &\approx& \Bigl \{1 + \int_0^1 d\lambda \ e^{\lambda\alpha A_0}
(\alpha \Delta A) e^{-\lambda\alpha A_0} \Bigr\}\, e^{\alpha A_0}  \nonumber\\
&=& \Bigl\{ 1 + \sum_{n=0}^\infty {\alpha^{n+1} \over (n+1)!} 
[A_0,\ldots [A_0, \Delta A] \ldots ] \Bigr\}\, e^{\alpha A_0}  \nonumber\\ 
\label{A12}
\end{eqnarray}
where the number of operators $A_0$ in the multiple commutator is $n$. 
Analogously, we expand $e^{\alpha B}=e^{\alpha B_0 + \alpha \Delta B}$ up to
first order in $\Delta B$ according to 
\begin{eqnarray}
e^{\alpha B} &\approx& e^{\alpha B_0}\, \Bigl \{1 + \int_0^1 d\lambda 
\ e^{-\lambda\alpha B_0} (\alpha \Delta B) e^{\lambda\alpha B_0} \Bigr\}
\nonumber\\
&=& e^{\alpha B_0}\, \Bigl\{ 1 + \sum_{n=0}^\infty {\alpha^{n+1} \over (n+1)!} 
[\ldots[\Delta B,B_0],\ldots B_0] \Bigr\}  \nonumber\\ 
\label{A13}
\end{eqnarray}
where the number of operators $B_0$ in the multiple commutator is $n$. 

Next, we evaluate the multiple commutators explicitly and resum the series' 
over $n$ by the hyperbolic functions. For convenience we define the 
vectors
\begin{eqnarray}
{\bf S} &=& {\bf b}_1 - i(g/2\gamma\Gamma) {\bf b} ,  \label{A14}  \\
{\bf S}^* &=& {\bf b}_1 + i(g/2\gamma\Gamma^*) {\bf b}  \label{A15} 
\end{eqnarray}
so that $\Delta A=-\Gamma {\bf S}{\bf r}$ and $\Delta B=-\Gamma^* {\bf S}^*
{\bf r}$. The rotation frequency ${\bf \Omega}=\Omega {\bf e}_z$ is 
incorporated via the wave vector ${\bf k}$, which is defined by (\ref{B04})
and which represents the uniform rotation of the normal-fluid component. 
Eventually, 
we obtain the operators
\begin{eqnarray}
e^{\alpha A} &=& \bigl\{ 1 - \alpha \Gamma {\bf S} {\bf r}  
-(\alpha\Gamma)^2 ({\bf e}_z {\bf S}) ({\bf e}_z (\bbox{\nabla}-i{\bf k})) 
\nonumber\\
&&\hskip-1cm +{\textstyle{1\over 2}} [\mbox{ch}(2\Gamma\alpha/\ell^2) -1]\,
\ell^4\, ({\bf e}_z\times ({\bf e}_z\times {\bf S})) (\bbox{\nabla}-i{\bf k}) 
\nonumber\\
&&\hskip-1cm -{\textstyle{i\over 2}} [\mbox{sh}(2\Gamma\alpha/\ell^2) - 
(2\Gamma\alpha/\ell^2) ]\, \ell^4\, ({\bf e}_z\times {\bf S}) 
(\bbox{\nabla}-i{\bf k}) \bigr\}\, e^{\alpha A_0}  \nonumber\\
\label{A16}
\end{eqnarray}
and
\begin{eqnarray}
e^{\alpha B} &=& e^{\alpha B_0}\, \bigl\{ 1 - \alpha\Gamma^* {\bf S}^* {\bf r} 
+(\alpha\Gamma^*)^2 ({\bf e}_z {\bf S}^*) ({\bf e}_z (\bbox{\nabla}-i{\bf k})) 
\nonumber\\
&&\hskip-1cm -{\textstyle{1\over 2}} [\mbox{ch}(2\Gamma^*\alpha/\ell^2) -1]\,
\ell^4\, ({\bf e}_z\times ({\bf e}_z\times {\bf S}^*)) (\bbox{\nabla}-
i{\bf k}) \nonumber\\
&&\hskip-1cm -{\textstyle{i\over 2}} [\mbox{sh}(2\Gamma^*\alpha/\ell^2) - 
(2\Gamma^*\alpha/\ell^2) ]\, \ell^4\, ({\bf e}_z\times {\bf S}^*) 
(\bbox{\nabla}-i{\bf k}) \bigr\}   \nonumber\\
\label{A17}
\end{eqnarray}
where $\ell=(\hbar/2m_4\Omega)^{1/2}$. Now, we insert these operators into 
the formula for the Green's function (\ref{A01}) and obtain
\begin{equation}
G({\bf r},{\bf r}^\prime) = 4\Gamma^\prime \int_0^\infty d\alpha 
\ \{\cdots\}\, e^{\alpha A_0} \, e^{\alpha B_0} \, \{\cdots\} 
\, \delta({\bf r}-{\bf r}^\prime)  \ . 
\label{A18}
\end{equation}
where the first curved brackets $\{\cdots\}$ are identified by the 
expressions in the curved brackets in (\ref{A16}) and the second curved 
brackets $\{\cdots\}$ are identified by the expressions in the curved 
brackets in (\ref{A17}). Because of $[A_0,B_0]=0$ we find 
\begin{equation}
e^{\alpha A_0} \, e^{\alpha B_0} = e^{\alpha (A_0+B_0)} =
e^{-\alpha 2\Gamma^\prime [a_1-(\bbox{\nabla}-i{\bf k})^2]} \ . 
\label{A19}
\end{equation}
In (\ref{A18}) in the second curved brackets the operator $\bbox{\nabla}$
acts directly on the delta function so that we can use the identity 
\begin{equation}
\bbox{\nabla}\, \delta({\bf r}-{\bf r}^\prime) =
-\bbox{\nabla}^\prime \, \delta({\bf r}-{\bf r}^\prime)  \ . 
\label{A20}
\end{equation}
Analogous identities are valid for ${\bf r}$ and ${\bf k}$. Thus, we obtain
the Green's function
\begin{eqnarray}
G({\bf r},{\bf r}^\prime) &=& 4\Gamma^\prime \int_0^\infty d\alpha 
\ \{\cdots\}\, \{\cdots\}^\prime \nonumber\\
&&\hskip0.7cm \times\, e^{-\alpha 2\Gamma^\prime[a_1- (\bbox{\nabla}-
i{\bf k})^2]} \, \delta({\bf r}-{\bf r}^\prime)  
\label{A21}
\end{eqnarray}
where in the second curved brackets $\{\cdots\}^\prime$ the operators 
$\bbox{\nabla}$, ${\bf r}$, and ${\bf k}$ are replaced by 
$-\bbox{\nabla}^\prime$, ${\bf r}^\prime$, and ${\bf k}^\prime$, respectively.
(${\bf k}^\prime$ is defined by (\ref{B04}) with ${\bf r}$ replaced by 
${\bf r}^\prime$.) We evaluate the product $\{\cdots\}\{\cdots\}^\prime$
up to the terms first order in ${\bf S}$ and ${\bf S}^*$. Furthermore, we
substitute $2\Gamma^\prime\alpha \rightarrow \alpha$ in the integral. Then,
from (\ref{A21}) we obtain the Green's function
\begin{eqnarray}
G({\bf r},{\bf r}^\prime) &=& 2 \int_0^\infty d\alpha 
\, \Bigl\{ 1 - {\alpha\Gamma\over 2\Gamma^\prime} {\bf S}{\bf r} 
- {\alpha\Gamma^*\over 2\Gamma^\prime} {\bf S}^*{\bf r}^\prime \nonumber\\ 
&&- \Bigl({\alpha\Gamma\over 2\Gamma^\prime}\Bigr)^2 ({\bf e}_z {\bf S})
({\bf e}_z (\bbox{\nabla}-i{\bf k}))  \nonumber\\
&&- \Bigl({\alpha\Gamma^*\over 2\Gamma^\prime}\Bigr)^2 ({\bf e}_z {\bf S}^*)
({\bf e}_z (\bbox{\nabla}^\prime+i{\bf k}^\prime))  \nonumber\\ 
&&\hskip-1cm +{1\over 2} \Bigl[ \mbox{ch} \Bigl({\Gamma\,\alpha\over
\Gamma^\prime\ell^2} \Bigr) -1 \Bigr] \ell^4 ({\bf e}_z\times 
({\bf e}_z\times {\bf S})) (\bbox{\nabla}-i{\bf k})  \nonumber\\
&&\hskip-1cm +{1\over 2} \Bigl[ \mbox{ch} \Bigl({\Gamma^*\alpha\over
\Gamma^\prime\ell^2} \Bigr) -1 \Bigr] \ell^4 ({\bf e}_z\times 
({\bf e}_z\times {\bf S}^*)) (\bbox{\nabla}^\prime+i{\bf k}^\prime)  
\nonumber\\
&&\hskip-1cm -{i\over 2} \Bigl[ \mbox{sh} \Bigl({\Gamma\,\alpha\over
\Gamma^\prime\ell^2}\Bigr) - \Bigl({\Gamma\,\alpha\over\Gamma^\prime\ell^2}
\Bigr) \Bigr] \ell^4 ({\bf e}_z\times {\bf S}) (\bbox{\nabla}-i{\bf k}) 
\nonumber\\
&&\hskip-1cm +{i\over 2} \Bigl[ \mbox{sh} \Bigl({\Gamma^*\alpha\over
\Gamma^\prime\ell^2}\Bigr) - \Bigl({\Gamma^*\alpha\over\Gamma^\prime\ell^2}
\Bigr) \Bigr] \ell^4 ({\bf e}_z\times {\bf S}^*) (\bbox{\nabla}^\prime+
i{\bf k}^\prime)\Bigr\} \nonumber\\
&&\hskip0.5cm \times\, e^{-\alpha [a_1- (\bbox{\nabla}-i{\bf k})^2]} 
\, \delta({\bf r}-{\bf r}^\prime)  \ .  
\label{A22}
\end{eqnarray}
Since ${\bf k}$ defined in (\ref{B04}) has the same form as the vector 
potential of a homogeneous magnetic field, the eigenfunctions and eigenvalues
of the operator $(\bbox{\nabla}-i{\bf k})^2$ are the Landau levels where
$\ell=(\hbar/2m_4 \Omega)^{1/2}$ is the ``magnetic length''. The factor in
the last line of (\ref{A22}) can be evaluated by representing the delta 
function in terms of the Landau-level eigenfunctions. Replacing the 
exponential operator by the related eigenvalues and performing the sum over
all eigenvalues, we obtain eventually
\begin{eqnarray}
e^{-\alpha [a_1- (\bbox{\nabla}-i{\bf k})^2]}\, \delta({\bf r}-{\bf r}^\prime)
&=& \nonumber\\
&&\hskip-3.5cm = \, e^{-i{\bf e}_z({\bf r}\times{\bf r}^\prime)/2\ell^2} \,
{1\over (4\pi\alpha)^{d/2}} \, {\alpha/\ell^2 \over \mbox{sh}(\alpha/\ell^2)}
\nonumber\\
&&\hskip-3.0cm \times\ \exp\Bigl( -\alpha a_1 -{\rho^2\over 4\alpha}
{\alpha/\ell^2 \over \mbox{th}(\alpha/\ell^2)} -{\zeta^2\over 4\alpha} \Bigr)
\label{A23}
\end{eqnarray}
where $\rho^2=(x-x^\prime)^2 + (y-y^\prime)^2$ and $\zeta^2=({\bf r}-
{\bf r}^\prime)^2 - \rho^2$. 

In the next step we apply the differential operators $\bbox{\nabla}$ and
$\bbox{\nabla}^\prime$ in the curved brackets of (\ref{A22}) onto the factor 
(\ref{A23}). We insert (\ref{A14}) and (\ref{A15}) for ${\bf S}$ and 
${\bf S}^*$. Since the dynamic parameters of model {\it F\,} occur only in
dimensionless ratios, from now on we write the formulas in terms of the
dimensionless renormalized parameters \cite{D1} $w=\Gamma/\lambda$, 
$w^*=\Gamma^*/\lambda$, and $F=g/\lambda$. Then, after some manipulations
of the terms in the curved brackets in (\ref{A22}) by using the addition
theorems of the hyperbolic functions and by using $\Gamma/\Gamma^\prime=
w/w^\prime=1+iw^{\prime\prime}/w^\prime$ and $\Gamma^*/\Gamma^\prime=
w^*/w^\prime=1-iw^{\prime\prime}/w^\prime$, we obtain the following result
for the Green's function:
\begin{eqnarray}
G({\bf r},{\bf r}^\prime) &=& 
e^{-i{\bf e}_z({\bf r}\times{\bf r}^\prime)/2\ell^2} \nonumber\\
&\times& \, {2\over (4\pi)^{d/2}} \int_0^\infty {d\alpha \over \alpha^{d/2}} 
\, \Bigl\{ 1 - \alpha{\bf b}_1 ({\bf r}+{\bf r}^\prime)/2 \nonumber\\
&&+\, i\, ({\bf e}_z\cdot ({\bf r}-{\bf r}^\prime)) \,\alpha {F\over 8\gamma 
w^\prime} ({\bf e}_z \cdot {\bf b})  \nonumber\\
&&+\, i\, ({\bf r}-{\bf r}^\prime) {\ell^2\over 2} \Bigl[ \Bigl( 1 -
{\alpha/\ell^2 \over \mbox{th}(\alpha/\ell^2)} \Bigr) ({\bf e}_z\times
{\bf b}_1) \nonumber\\
&&-{1\over \mbox{sh}(\alpha/\ell^2)}\, {F\over 4\gamma} \, {2w^\prime\over
{w^\prime}^2 + {w^{\prime\prime}}^2} \nonumber\\
&&\times\, \Bigl( \Bigl[ \mbox{ch}\Bigl({\alpha\over
\ell^2}\Bigr) - \cos \Bigl( {w^{\prime\prime}\alpha\over w^\prime \ell^2 }
\Bigr) \Bigr] \,({\bf e}_z\times ({\bf e}_z\times {\bf b})) \nonumber\\
&&+ \Bigl[ {w^{\prime\prime}\over w^\prime} \mbox{sh}\Bigl({\alpha\over
\ell^2}\Bigr) - \sin \Bigl( {w^{\prime\prime}\alpha\over w^\prime \ell^2 }
\Bigr) \Bigr] \, ({\bf e}_z \times {\bf b}) \Bigr) \Bigr] \Bigr\}
\nonumber\\
&\times& {\alpha/\ell^2 \over \mbox{sh}(\alpha/\ell^2)} \, 
\exp\Bigl( -\alpha a_1 -{\rho^2\over 4\alpha} {\alpha/\ell^2 \over 
\mbox{th}(\alpha/\ell^2)} -{\zeta^2\over 4\alpha} \Bigr)  \ .  \nonumber\\
\label{A24}
\end{eqnarray}
Finally, we use the formula $e^x\approx \{1+x\}$ for small $x$ to combine the
terms in the curved brackets into an exponential function. Thus, eventually
we obtain
\begin{eqnarray}
G({\bf r},{\bf r}^\prime) &=& 
e^{-i{\bf e}_z({\bf r}\times{\bf r}^\prime)/2\ell^2} 
\, {2\over (4\pi)^{d/2}} \int_0^\infty {d\alpha \over \alpha^{d/2}} 
\, {\alpha/\ell^2 \over \mbox{sh}(\alpha/\ell^2)} \nonumber\\
&&\hskip-1cm \times \exp\Bigl( -\alpha \bar r_1 -{\rho^2\over 4\alpha} 
{\alpha/\ell^2 \over \mbox{th}(\alpha/\ell^2)} -{\zeta^2\over 4\alpha} \Bigr)  
\, e^{i{\bf K}({\bf r}-{\bf r}^\prime)} 
\label{A25}
\end{eqnarray}
where
\begin{equation}
\bar r_1 = a_1 + {\bf b}_1 ({\bf r}+{\bf r}^\prime)/2 
\label{A26}
\end{equation}
and
\begin{eqnarray}
{\bf K} &=& \alpha {F\over 8\gamma w^\prime} \,{\bf e}_z ({\bf e}_z \cdot 
{\bf b})  \nonumber\\
&+& {\ell^2\over 2} \Bigl[ \Bigl( 1 - {\alpha/\ell^2 \over \mbox{th}
(\alpha/\ell^2)} \Bigr) ({\bf e}_z\times {\bf b}_1) \nonumber\\
&&-{1\over \mbox{sh}(\alpha/\ell^2)}\, {F\over 4\gamma} \, {2w^\prime\over
{w^\prime}^2 + {w^{\prime\prime}}^2} \nonumber\\
&&\times\, \Bigl( \Bigl[ \mbox{ch}\Bigl({\alpha\over
\ell^2}\Bigr) - \cos \Bigl( {w^{\prime\prime}\alpha\over w^\prime \ell^2 }
\Bigr) \Bigr] \,({\bf e}_z\times ({\bf e}_z\times {\bf b})) \nonumber\\
&&+ \Bigl[ {w^{\prime\prime}\over w^\prime} \mbox{sh}\Bigl({\alpha\over
\ell^2}\Bigr) - \sin \Bigl( {w^{\prime\prime}\alpha\over w^\prime \ell^2 }
\Bigr) \Bigr] \, ({\bf e}_z \times {\bf b}) \Bigr) \Bigr]  \ . 
\label{A27}
\end{eqnarray}
$\rho^2$ and $\zeta^2$ are defined below (\ref{A23}). Eq.\ (\ref{A25}) 
together with (\ref{A26}) and (\ref{A27}) is the final result for the 
Green's function $G({\bf r},{\bf r}^\prime)$, which is valid for infinitesimal
gradients ${\bf b}_1=\bbox{\nabla} r_1$ and ${\bf b}=\bbox{\nabla}(\Delta r)$ 
up to first order in ${\bf b}_1$ and ${\bf b}$. For ${\bf K}={\bf 0}$ and 
$\bar r_1= r_1=\mbox{const.}$ Eq.\ (\ref{A25}) represents the Green's function 
for rotating $^4$He in thermal equilibrium. 

The wave vector ${\bf K}$ defined in (\ref{A27}) is the sum of the two terms 
of different nature which are orthogonal to each other. The first term in 
(\ref{A27}) represents the contribution parallel to the rotation axis and
does not depend on the rotation frequency $\Omega$. The second term, which is
given by the remaining terms in (\ref{A27}), is in the $xy$ plane 
perpendicular to the rotation axis and depends on $\Omega$ via the 
characteristic length $\ell$. In the limit $\Omega\rightarrow 0$ and $\ell
\rightarrow\infty$ the wave vector reduces to ${\bf K}\rightarrow\alpha
(F/8\gamma w^\prime){\bf b}$. This wave vector is well known from the theory 
of heat transport in nonrotating $^4$He (see e.g.\ (A28) in the second 
paper of Ref.\ \onlinecite{H1}). 

While the scalar and vector products involving ${\bf e}_z$ are defined in 
$d=3$ dimensions, our formula for $G({\bf r},{\bf r}^\prime)$ is valid for 
arbitrary $d$ dimensions. The scalar and vector products in (\ref{A25}) and 
(\ref{A27}) must be replaced by appropriate generalized expressions. This, 
however, is straightforward: $-{\bf e}_z\times ({\bf e}_z\times {\bf b})$ is 
replaced by the projection of ${\bf b}$ onto the $xy$ plane and ${\bf e}_z 
({\bf e}_z\cdot {\bf b})$ is replaced by the projection of ${\bf b}$ onto the 
subspace perpendicular to the $xy$ plane. Furthermore, ${\bf e}_z\times 
{\bf b}_1$ and ${\bf e}_z\times {\bf b}$ are replaced by the projections of 
${\bf b}_1$ and ${\bf b}$ onto the $xy$ plane combined with a rotation of 
angle $\pi/2$. 

The density $n_{\rm s}$ and the superfluid current ${\bf J}_{\rm s}$ are 
obtained by inserting the Green's function (\ref{A25}) into (\ref{B12}) and
(\ref{B19}), respectively. We find
\begin{equation}
n_{\rm s}= {2\over (4\pi)^{d/2}} \int_0^\infty {d\alpha \over \alpha^{d/2}} 
\, {\alpha/\ell^2 \over \mbox{sh}(\alpha/\ell^2)} \, e^{-\alpha r_1} \ ,
\label{A28}
\end{equation}
and
\begin{equation}
{\bf J}_{\rm s}= {2\over (4\pi)^{d/2}} \int_0^\infty {d\alpha \over 
\alpha^{d/2}} \, {\alpha/\ell^2 \over \mbox{sh}(\alpha/\ell^2)} \, 
{\bf K}\, e^{-\alpha r_1} 
\label{A29}
\end{equation}
where ${\bf K}$ is defined in (\ref{A27}). We replace
\begin{equation}
{1\over (4\pi)^{d/2}} = - {1\over\epsilon} A_d {1\over \Gamma(-1+\epsilon/2)}
\label{A30}
\end{equation}
where $A_d$ is the geometrical factor which occurs in the renormalization
equations (\ref{B22}), (\ref{B31}), and (\ref{B32}). Furthermore, we 
substitute $\alpha/\ell^2\rightarrow v$. Thus, eventually we obtain the
formulas (\ref{B25}) and (\ref{B37}) with the integrals (\ref{B26}) and 
(\ref{B38})-(\ref{B40}).


\begin{thebibliography}{10}
\bibitem{DO} R. J. Donnelly, {\it Quantized Vortices in Helium II}
(Cambridge Univ. Press, Cambridge 1991).
\bibitem{HH} B. I. Halperin, P. C. Hohenberg, and E. D. Siggia, Phys. Rev. 
Lett. {\bf 32}, 1289 (1974); Phys. Rev. B {\bf 13}, 1299 (1976).
\bibitem{F1} R. P. Feynman, in: Progress in Low Temperature Physics {\bf 1},
edited by C. J. Gorter (North Holland, Amsterdam 1955), Chap. 2.
\bibitem{H1} R. Haussmann, J. Low Temp. Phys. {\bf 114}, 1 (1999) and 
submitted to Phys. Rev. B (1998).
\bibitem{HV} H. E. Hall and W. F. Vinen, Proc. R. Soc. A {\bf 238}, 204 and 
215 (1956).
\bibitem{H2} R. Haussmann, Z. Phys. B {\bf 87}, 247 (1992).
\bibitem{GM} C. J. Gorter and J. H. Mellink, Physica {\bf 15}, 285 (1949).
\bibitem{V1} W. F. Vinen, Proc. R. Soc. A {\bf 240}, 114 and 128 (1957);
{\bf 242}, 493 (1957); {\bf 243}, 400 (1958). 
\bibitem{A0} G. Ahlers, Phys. Rev. {\bf 171}, 275 (1968).
\bibitem{SL} H. A. Snyder and D. M. Lynekin, Phys. Rev. {\bf 147}, 131 (1966).
\bibitem{L1} P. Lucas, J. Phys. C {\bf 3}, 1180 (1970).
\bibitem{MS} P. Mathieu, A. Serra, and Y. Simon, Phys. Rev. B {\bf 14}, 3753
(1976). 
\bibitem{D1} V. Dohm, Z. Phys. B {\bf 60}, 61 (1985); Z. Phys. B {\bf 61},
193 (1985); Phys. Rev. B {\bf 44}, 2697 (1991). 
\bibitem{SD} R. Schloms and V. Dohm, Nucl. Phys. B {\bf 328}, 639 (1989).
\bibitem{TA} W. Y. Tam and G. Ahlers, Phys. Rev. B {\bf 32}, 5932 (1985).
\bibitem{LS} J. A. Lipa, D. R. Swanson, J. A. Nissen, T. C. P. Chui, and
U. E. Israelsson, Phys. Rev. Lett. {\bf 76}, 944 (1996).
\bibitem{HD2} R. Haussmann and V. Dohm, Phys. Rev. B {\bf 46}, 6361 (1992).
\bibitem{BA} H. Baddar, G. Ahlers, K. Kuehn, and H. Fu, private communication
(1998).
\end{thebibliography}
\end{document}